\documentclass[aps,onecolumn,groupedaddress,amsmath,amssymb,nofootinbib]{revtex4}
\usepackage{amsmath, dsfont}
\usepackage{amssymb}
\usepackage{epsfig}
\usepackage{calc}
\usepackage{amsfonts}
\usepackage{graphicx}
\usepackage[ansinew]{inputenc}
\usepackage{multirow}
\usepackage{subcaption}
\usepackage{tensor}
\DeclareMathOperator\arctanh{arctanh}
\usepackage[dvipsnames]{xcolor}

\begin{document}
\setcounter{page}{1}
\title[]{Conformally coupled theories and their deformed compact objects: \\ from black holes, radiating spacetimes to eternal wormholes}
\author{Eugeny Babichev$^{1}$, Christos Charmousis$^{1}$, Mokhtar Hassaine$^{2}$ and  Nicolas Lecoeur$^{1}$\\
    $^{1}$ Universit\'e Paris-Saclay, CNRS/IN2P3, IJCLab, 91405 Orsay, France \\
    $^{2}$ Instituto de Matem\'{a}tica y F\'{\i}sica, Universidad de
    Talca, Casilla 747, Talca, Chile}

\begin{abstract}
We study a higher order conformally coupled scalar tensor theory
endowed with a covariant geometric constraint relating the scalar
curvature with the Gauss-Bonnet scalar. It is a particular 
Horndeski theory including a canonical kinetic term but without shift or
parity symmetry for the scalar. The theory also stems from a Kaluza-Klein reduction of a well defined 
higher dimensional metric theory. Properties of an asymptotically
flat spherically symmetric black hole are analyzed, and new slowly
rotating and radiating extensions are found. Through
disformal transformations of the static configurations, gravitating
monopole-like solutions and eternal wormholes are presented. The
latter are shown to extract from spacetime possible naked
singularities, yielding completely regular and asymptotically flat
spacetimes.
\end{abstract}

\maketitle

\tableofcontents \thispagestyle{empty}

\section{Introduction}
In recent years, scientific interest and research in black holes, neutron stars and other
more exotic compact objects, such as wormholes \cite{Morris:1988cz},
has increased considerably. This is largely due to the plethora of
recent astrophysical observations \cite{LIGOScientific:2017vwq,
LIGOScientific:2020zkf, LIGOScientific:2021qlt} which confirm or
re-affirm, the existence of compact objects as well as their
defining properties. These observations are in their vast majority
in accordance with General Relativity (GR) at their current
accuracy. Certain unexpected results do emerge however, questioning
certain standard expectations from GR. For example, the recent
observation of the compact object merger GW190814
\cite{LIGOScientific:2020zkf} where the secondary compact object has
a mass of $2.59_{-0.09}^{+0.08}~M_\odot$, placing it in the mass gap
in-between neutron stars and black holes for GR. From classical GR
results such as Buchdahl limit on compacity, such a compact object
of astrophysical origin could be explained only as a neutron star
with an unexpectedly stiff (or exotic) EOS (quite incompatible with
GW170817), a neutron star with a too 
rapid rotation, or a black hole with a small mass whose origin is
difficult to explain (for a discussion see \cite{Charmousis:2021npl}
and references within).

It is clear that we are entering a novel era in gravitational
observations, and technological/observational advances in the near
future will definitely bring to light new aspects of gravitational physics, some of which probably not anticipated, that we will still have to comprehend.
We are presented therefore with quite a challenge in gravitational theory
with the need to extend our understanding concerning the existence
and properties of compact objects as solutions of GR or other
theories of gravity. It is also important to emphasize that although
most current observational data are in agreement with the theory of
GR, this should in no way prevent us from exploiting alternative
gravity theories as they provide a measurable ruler of departure
from classical relativity theory. In this perspective, it is certain
that the emergence of new gravitational solutions (associated with
modified theories) will enrich our understanding of recent and
future observations. Therefore, it is crucial 
to search for
modifications of GR and to explore new promising theoretical
possibilities in theories of gravity. In order to carry out this
project, we must specify our modified theories of gravity so that
they are physically acceptable while also ensuring the existence of
analytical solutions, which are an important condition 
for making accurate comparison of GR and its modifications using observations.

Modifications of gravity can be realized with increasingly complex
formulations but, in the present case, we will be restricting
ourselves to scalar-tensor theories which are the simplest, working,
robust prototype of modified gravity theories with a single
additional degree of freedom. They also appear as a limit of most
modified gravity theories however complex their
nature.
In recent years, higher order scalar
tensor theories (with second-order field equations) have been rediscovered, and intensively studied highlighting the precursor
work of Horndeski \cite{Horndeski:1974wa} from the seventies. For latter convenience, we specify the Horndeski action which is
nothing  but the most general (single) scalar-tensor theory with second order equations of motion,
\begin{eqnarray}
S = \int \mathrm{d}^4x\sqrt{-g}\, \Big\lbrace \mathcal{L}_2+\mathcal{L}_3+\mathcal{L}_4+\mathcal{L}_5\Big\rbrace, \label{eqaction}
\end{eqnarray}
with
\begin{eqnarray*}
&&\mathcal{L}_2 = G_2, \quad \mathcal{L}_3=-G_3\Box\phi,\quad \mathcal{L}_4 =  G_4R+G_{4X}\left[\left(\Box\phi\right)^2 -\left(\phi_{\mu\nu}\right)^2 \right], \\
&&\mathcal{L}_5 = G_5\,G_{\mu\nu}\phi^{\mu\nu} -
\frac{1}{6}G_{5X}\left(\left(\Box\phi\right)^3 -
3\Box\phi\left(\phi_{\mu\nu}\right)^2
+2\phi_{\mu\nu}\phi^{\nu\rho}\phi{^\mu_\rho}\right),
\label{eqaction1}
\end{eqnarray*}
where $\phi_\mu = \nabla_\mu\phi$, $\phi_{\mu\nu}=\nabla_\mu\nabla_\nu\phi$, and the $G_k$'s are arbitrary functions of $\phi$ and of the standard kinetic term
$X=-\phi_\mu\phi^\mu/2$ parametrising the Horndeski theory.

Sectors of the Horndeski
theory and beyond have been exploited in the current literature  (see \cite{Babichev:2013cya, 
BenAchour:2019fdf, Babichev:2017guv} and references therein) providing explicit compact object solutions and related results. As it turns out, the theories 
which allow analytic construction of solutions are, mostly restricted to a shift-symmetric and parity-preserving scalar field\footnote{These are Horndeski theories that are invariant under the constant shift of the scalar field $\phi\to\phi+\mbox{cst}$ and parity symmetry $\phi \to -\phi$}. 
The shift-symmetry of the scalar field yields a Noether conserved
current which proves extremely useful for integrating the equations of motion. The lesson to be learned from these examples is that symmetries
underlying the action of the scalar tensor theories (\ref{eqaction}) are key in obtaining workable analytic solutions.
From this observation, it is natural to focus in the classes of Horndeski theories 
possessing symmetries simplifying the equations of motion. 
Such a symmetry could also be the conformal invariance of the equation of motion of the scalar field.
The advantage of the latter is the existence of a covariant purely geometric constraint which does not involve the scalar field.
This idea is not new and finds its origin in the first counter-example to
the no-hair theorem with the discovery of the so-called BBMB black hole \cite{Bocharova:1970skc, Bekenstein:1974sf} which corresponds to
a static solution of the Einstein equations with a conformally coupled scalar field in four dimensions.{\footnote{It is interesting to note that the
extension of the BBMB solution in higher dimensions leads to singular metrics
\cite{Klimcik:1993cia}}
In this case, the purely geometric equation which allows the integration of the equations of motion is the
vanishing Ricci scalar, $R=0$. 
In presence of a cosmological constant with a self-interacting potential, this constraint is modified to $R=\mbox{cst}$, while conformal invariance for the scalar is not spoilt. As a result analytic black hole solutions of de Sitter and anti de Sitter asymptotics were found in \cite{Martinez:2002ru, Martinez:2005di}. 
Quite recently this approach was nicely extended to the most general (higher order) Horndeski action with a conformally-invariant scalar field equation~\cite{Fernandes:2021dsb},}
\begin{eqnarray}
S = \int \mathrm{d}^4x\frac{\sqrt{-g}}{16\pi}\left\lbrace R-2\lambda
e^{4\phi}  - \beta
\mathrm{e}^{2\phi}\left(R+6\left(\nabla\phi\right)^2\right) -
\alpha\left[\phi\mathcal{G}-4G^{\mu\nu}\phi_\mu\phi_\nu-4\Box\phi\left(\nabla\phi\right)^2-2\left(\nabla\phi\right)^4\right]\right\rbrace,
\label{eq:action}
\end{eqnarray}
and, {\it cerise sur le g\^ateau}, this action belongs to a
non-shift symmmetric Horndeski class (\ref{eqaction}) without parity symmetry.
Indeed all the Horndeski   coupling functions are present taking the form,
\begin{eqnarray}
G_2= -2\lambda e^{4\phi}+12\beta e^{2\phi} X+8\alpha X^2,\quad
G_3=8\alpha X,\quad G_4=1-\beta e^{2\phi}+4\alpha X,\quad
G_5=4\alpha \ln \left\lvert X\right\rvert. \label{couplingfcts}
\end{eqnarray}
Here $\alpha$, $\beta$ and $\lambda$ are constant parameters and  $\mathcal{G}=R^2-4R_{\mu\nu}R^{\mu\nu}+R_{\mu\nu\rho\sigma}R^{\mu\nu\rho\sigma}$
is the Gauss-Bonnet scalar, while a cosmological constant may also be added to the action (\ref{eq:action}). 
The particularity of the construction in \cite{Fernandes:2021dsb}
however, is that the trace of the metric equations together with the
scalar field equation associated to the action (\ref{eq:action})
combine to give a purely geometric four-dimensional equation,
\begin{equation}
R+\frac{\alpha}{2}\mathcal{G}=0.
\label{eq:geom}
\end{equation}
With the help of this geometric constraint, two analytic static solutions, with nontrivial scalar fields, were presented in~\cite{Fernandes:2021dsb}, for $\beta\neq 0$. In fact, each of them exists for a precise tuning between the coupling constants $\alpha$, $\beta$ and $\lambda$ in action (\ref{eq:action}), so the associated theories are distinct. We will focus on one of these solutions and its corresponding theory, which presents the attractive feature of both a canonical kinetic term and a well-defined scalar field in the whole spacetime (minus the origin). Last but not least, the latter solution also has a higher dimensional origin. 
Indeed it is interesting to note that the above action~(\ref{couplingfcts}) can be approached from an alternative route involving the Kaluza-Klein compactification of $D-$dimensional Einstein-Gauss-Bonnet theory \cite{Kiritsis}. There it was shown that starting from a $D>4$ dimensional solution of Lovelock gravity with a non trivial horizon \cite{Dotti:2005rc,Bogdanos}, one can construct a scalar tensor black hole solution in four dimensions~\cite{Kiritsis}. These solutions, due to their higher dimensional origin, do not have a standard four dimensional Newtonian  mass term. Crucially however, upon taking a singular limit (as first considered by \cite{Glavan}), action (\ref{eq:action}) and the latter solution from \cite{Fernandes:2021dsb}, can be obtained from~\cite{Kiritsis} with a standard four dimensional mass term.

We thus provide a
detailed analysis of this solution in the first part of the present work, by studying the
nature of the singularities, depending on the sign of the coupling
constant $\alpha$. Indeed, we show that the case $\alpha>0$ is
well-behaved, with a spacetime defined in the whole region $r>0$,
and with a singularity at $r=0$ always hidden by a horizon, while
for $\alpha<0$, a naked singularity may appear. Then, starting from
the observation that the solutions of \cite{Fernandes:2021dsb} do
not reduce to flat spacetime, we seek non-trivial flat
spacetime solutions of the given theory. We present two classes of flat spacetime solutions with
a non trivial time-dependent scalar field. 
We furthermore extend the solution of~\cite{Fernandes:2021dsb} to find a slowly rotating black hole solution, as well as a
radiating/accreting Vaidya-like solution for this modified gravity theory. 

Another aspect that has been recently studied in the literature for
(beyond) Horndeski theories has to do with disformal transformations
of the metric, see Ref.~\cite{Zumalacarregui:2013pma}. Starting from a seed
solution given by a scalar field $\phi$ and a metric $g$ of a given
Horndeski theory, the deformed metric $\tilde{g}_{\mu\nu}=g_{\mu\nu}+D(\phi,
X)\partial_{\mu}\phi\partial_{\nu}\phi$ solves a beyond Horndeski theory, along with an unchanged scalar field. 
Disformal transformations are very useful in engineering solutions with highly non-trivial properties from simpler seed solutions.
In particular, in Ref.~\cite{Anson:2020trg}, disformal versions
of the Kerr spacetime with a regular scalar field were explicitly
constructed and analyzed starting from a stealth Kerr solution
\cite{Charmousis:2019vnf}. Such rotating black holes have
particular non-GR observational signatures \cite{Anson:2021yli}, which in the near future may be probed and contrasted with the Kerr solution.
Disformal transformations can also give rise to explicit
asymptotically flat wormhole solutions \cite{Bakopoulos:2021liw} (see also \cite{Faraoni:2021gdl}, \cite{Chatzifotis:2021hpg} and also \cite{Kanti:2011jz} for earlier works). We will exploit this direction in the second part of the paper to construct regular wormholes and regular monopole-like solutions.

In the next section, we will analyze the black holes in question,
portraying non trivial flat spacetime solutions as well as their
slowly rotating and Vaidya like counterparts. We will then in the
third section discuss ways to circumvent certain shortcomings of
the initial solution portraying in particular eternal wormhole
metrics as well as regular monopole-like solutions. We will conclude our
analysis discussing future prospects. For clarity, we will include slowly rotating and radiating extensions of other solutions to action (\ref{eq:action}), as well as the specific disformed theories of the latter action, in the appendix.

\section{A hairy black hole solution, its flat counterpart and generalizations}

\subsection{Black hole analysis}

The theory  under consideration (\ref{eq:action}) presents several
noteworthy properties. For a start, it is the most general scalar-tensor action with second-order equations of motion endowed with a conformally coupled scalar field
\cite{Fernandes:2021dsb}. Secondly, action (\ref{eq:action}) has
a higher dimensional origin from a purely metric theory, namely
Lovelock theory \cite{Lovelock:1971yv} (see
\cite{Charmousis:2008kc} for a review). In effect, the conformally coupled theory
can be also obtained in a two step fashion: from a consistent
Kaluza-Klein reduction of higher dimensional Lovelock theory
\cite{Kiritsis} where the dimension $D$ is a continuous parameter,
followed by a singular limit of $D\rightarrow 4$ as first considered
in \cite{Glavan}, and later applied in this context in
\cite{Lu:2020iav}. A third important fact is the presence, when $\beta\neq 0$, of  a
canonical kinetic term (obtained by a simple field redefinition $\Phi=\mathrm{exp}\left(\phi\right)$), and the absence of shift or parity symmetry. As a direct
consequence, this theory is not subject to the standard shift
symmetric Horndeski no hair theorem \cite{Hui:2012qt}, and hence it is not clear {\it{a priori}} which properties compact solutions of (\ref{eq:action})
may acquire. In fact, in a recent elegant paper
\cite{Fernandes:2021dsb}, the author finds distinct classes of
static solutions for the scalar tensor theory (\ref{eq:action})
with a particular tuning in between the coupling constants
$\lambda$, $\beta$ and $\alpha$ (see also \cite{Lu:2020iav}, \cite{Fernandes:2022zrq} and references within). Different cases, along with new solutions, will be discussed in the appendices, but in the main body of the
paper, we will focus on the unique solution of \cite{Fernandes:2021dsb} with both $\beta\neq 0$ and a scalar field with a logarithmic behavior which is well defined everywhere but the origin{\footnote{Note that to lowest order in $\alpha$,
this theory is nothing but the BBMB theory \cite{Bocharova:1970skc},
\cite{Bekenstein:1974sf} as can be easily verified by setting
$\Phi=\mathrm{exp}\left(\phi\right)$. However, the presently considered solution for the
scalar field is quite different, since it only blows up at
the origin and not at the horizon of the black hole, one of the
notorious setbacks of the BBMB solution. }}, and the couplings
satisfying the constraint $\lambda=\frac{\beta^2}{4\alpha}$. This
latter is given by
\begin{equation}
\label{eq:solutionA}
\mathrm{d}s^2=-f(r)\mathrm{d}t^2+\frac{\mathrm{d}r^2}{f(r)}+r^2\left(\mathrm{d}\theta^2+\sin^2\theta\,
\mathrm{d}\varphi^2\right),
\end{equation}
with
\begin{equation}
\label{eq:f}
f(r) =
1+\frac{r^2}{2\alpha}\left(1-\sqrt{1+8\alpha\left(\frac{M}{r^3}+\frac{\alpha}{r^4}\right)}\right)
\end{equation}
and
\begin{equation}
\label{eq:solutionAbis}
\phi =\phi(r)=\ln\left(\frac{\sqrt{-2\alpha/\beta}}{r}\right).
\end{equation}

The solution depends on a unique integration constant denoted by $M$
(and corresponding to the mass, as proven below), and exists provided the couplings
$\alpha$ and $\beta$ are of opposite sign. It is therefore a black
hole with secondary hair, as are most scalar-tensor black holes.
However, note that the scalar charge of this solution is not
trivial. Indeed, if we switch off the integration constant, $M=0$,
we do not end up with flat spacetime, rather a singular solution at
$r=0$ (with singularity covered by an event horizon for $\alpha>0$), and this is
essentially due to the additional $\alpha^2/r^4$ term under the
square root in (\ref{eq:f}). This latter term can be seen to be
related to the scalar charge of the black hole. Note in fact that at
$r=0$ the solution behaves as $f\left(r\right)\sim 1-\text{sign}\left(\alpha\right)\sqrt{2}+ O(r)$ which is
finite and certainly not equal to 1. This seemingly milder
singularity is a true curvature singularity at $r=0$, in agreement with the
logarithmically singular scalar field there. Therefore we see that
the canonical kinetic term does come at the expense of a singular
vacuum, therefore an essential question that will occupy us later on
in this section is the existence of a flat solution in this theory.

\begin{figure}
\begin{subfigure}{0.498\textwidth}
\includegraphics[width=\linewidth]{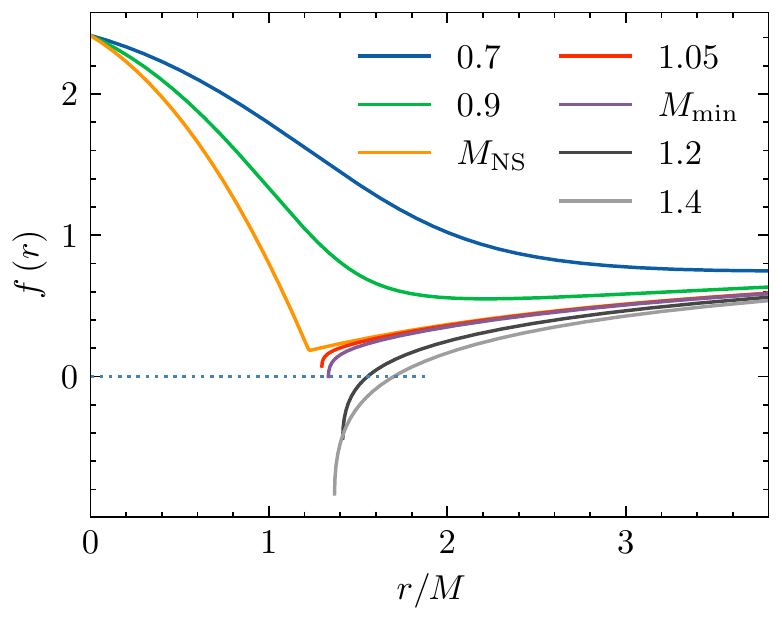}
\end{subfigure}
\begin{subfigure}{0.495\textwidth}
\includegraphics[width=\linewidth]{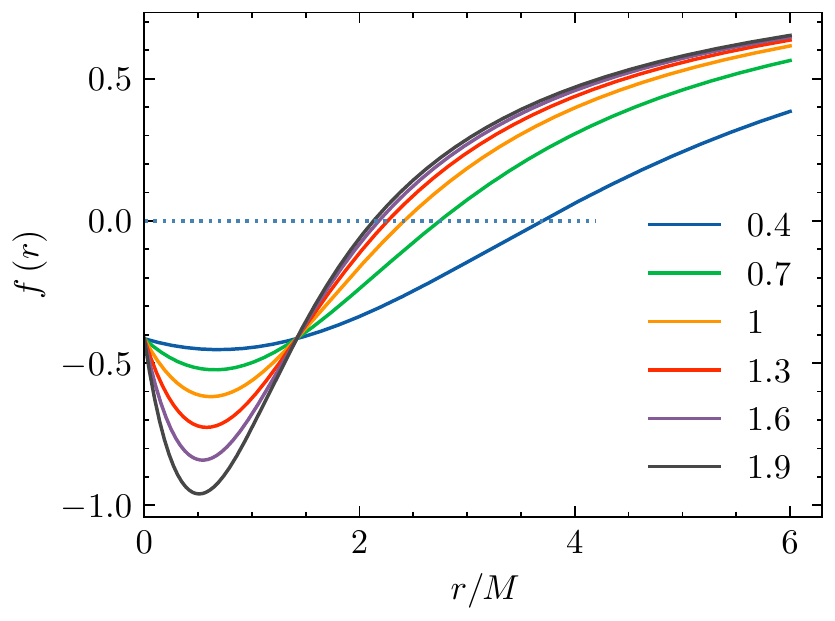}
\end{subfigure}
\caption{Metric function $f(r)$ for different values of $M/\sqrt{\left\lvert\alpha\right\rvert}$ for negative $\alpha$ (left plot) and positive $\alpha$ (right plot).
On the left panel, for $M\leq M_\text{NS}$, the upper curves correspond to the spacetime with a naked singularity at $r=0$.
For $M_0<M< M_\text{min}$, the spacetime has a naked singularity at $r=r_P$, while for $M\geq M_{\text{min}}$ the metric describes a black hole.
On the right panel, the spacetime admits a singularity at $r=0$, always covered by the horizon.}
\label{fig:f(r)fern}
\end{figure}

For the moment, let us pursue the study of the
spacetime~(\ref{eq:solutionA}). The spacetime for the solution
exhibits very distinct properties depending on the sign of the
coupling constant $\alpha$. For $\alpha<0$ (and hence $\beta>0$),
the standard kinetic term has the usual sign in the
action\footnote{This can be seen from the scalar field redefinition
$\Phi=\mathrm{exp}\left(\phi\right)$.}, and the coupling constant of
the potential term $\lambda=\frac{\beta^2}{4\alpha}<0$.
For convenience, we rewrite the spacetime (\ref{eq:solutionA}) for the choice $\alpha<0$ as follows,
\begin{equation}
f\left(r\right) =
1-\frac{r^2}{2\left\vert\alpha\right\vert}+\frac{\sqrt{P\left(r\right)}}{2\left\vert\alpha\right\vert},\quad
P\left(r\right)\equiv r^4-8\left\vert\alpha\right\vert Mr + 8
\left\vert\alpha\right\vert^2,\label{eq:fF}
\end{equation}
and we define the radius $r=r_P$ and the values $M_\text{NS}$ and $M_\text{min}$,
\begin{equation}
P\left(r_P\right)\equiv 0,\qquad
\frac{\left\vert\alpha\right\vert}{M_\text{NS}^2}\equiv
\frac{3}{4}\sqrt{\frac{3}{2}},\qquad
\frac{\left\vert\alpha\right\vert}{M_{\text{min}}^2}\equiv
\frac{8}{9}.
\label{M0}
\end{equation}
It is easy to see that for $0\leq M\leq M_\text{NS}$, the spacetime
admits a naked singularity at $r=0$, while if $M_\text{NS}<M<
M_\text{min}$, the naked singularity is brought forward to $r=r_P$.
Only for larger masses $M\geq M_\text{min}$ (as compared to the
coupling constant $\left\lvert\alpha\right\rvert$) does the spacetime describe a black
hole with a single event horizon at
$r_+=M+\sqrt{M^2-\left\vert\alpha\right\vert}$ covering the
singularity at $r=r_P$. Note that  for
$\alpha<0$ the event horizon has smaller
size compared to the standard Schwarzschild radius $r_\text{Sch}=2M$. In particular the minimal horizon size is $r_+^\text{min}=\sqrt{2\vert\alpha\vert}=\frac{4}{3}M_\text{min}$. The behavior of the metric function is illustrated in
Fig.~\ref{fig:f(r)fern} (left panel), where $f\left(r\right)$ is
shown for different $M/\sqrt{\left\lvert\alpha\right\rvert}$.

For $\alpha<0$, the lower bound on the mass
$M\geq M_\text{min}$ ensuring the existence of a black hole solution
implies an upper bound on the value of the
coupling parameter $\left\vert\alpha\right\vert$.
Indeed, following Ref.~\cite{Charmousis:2021npl}, one can obtain a constraint on $\alpha$ using data on observed (candidates of) black holes.
In the event GW$200115$, one component was certainly identified as a black hole of mass $M=5.7_{-2.1}^{+1.8}\, M_{\odot}$.
This gives a constraint
\begin{equation}
\left\vert\alpha\right\vert\lesssim
253_{-152}^{+184}\,\mbox{km}^2.
\end{equation}
If we include the events GW$170817$ and GW$190814$, we obtain
stronger constrains, $\left\vert\alpha\right\vert\lesssim
59\,\mbox{km}^2$, and $ \left\vert\alpha\right\vert\lesssim
52\,\mbox{km}^2,$ correspondingly, however the presence of a black
hole is only probable (but not certain) for these two events.

The case $\alpha>0$ is more straightforward to analyze since,
independently of the value for $\alpha$, the solution
(\ref{eq:solutionA}) describes a black hole for any mass $M$,
and with a unique horizon $r_+ = M+\sqrt{M^2+\alpha}$ covering the
singularity $r=0$. The horizon is now at $r_+>r_\text{Sch}=2M$. The behavior of the function $f(r)$ is illustrated
in Fig.~\ref{fig:f(r)fern}.

To conclude the discussion, we would like to mention, in the spirit
of~\cite{Charmousis:2021npl}, that if a Birkhoff-like uniqueness
theorem were valid for the
solution~(\ref{eq:solutionA}-\ref{eq:solutionAbis}), it would
inevitably lead to the constraint $\alpha<0$. Indeed, if the solution~(\ref{eq:solutionA}) were unique, any static and
spherically symmetric object of mass $M$ would create an exterior
gravitational field given by~(\ref{eq:solutionA}). If $\alpha>0$, this object would therefore be a black hole with horizon $r_+=M+\sqrt{M^2+\alpha}$, unless this event horizon is hidden below the surface of the object. An atomic nucleus has radius $R\sim
10^{-15}\,\mbox{m}$, and is not a black hole since it can be experimentally probed, therefore $r_+<R$, yielding
\begin{equation}
0<\alpha < R\left(R-2M\right) \sim 10^{-30}\,\mbox{m}^2,
\label{eq:constraint}
\end{equation}
essentially rendering $\alpha>0$ irrelevant.

\subsection{Black hole thermodynamics}

Let us now turn to the thermodynamic properties of the black holes
of ~(\ref{eq:action}). Since the theory in question descends from a
spin $2$ metric Lovelock theory, its thermodynamic aspects can be
quite intriguing \cite{Myers:1988ze}, \cite{Clunan:2004tb}. In
particular, one may ask whether the one-quarter area law of the
entropy is preserved or not. In order to give a clear answer we choose to use
the Euclidean approach for a general class of spherically symmetric
metrics parameterized as,
\begin{equation}
\label{euc}
\mathrm{d}s_{\mbox{\tiny{Eucl}}}^2=N(r)^2f(r)\mathrm{d}\tau^2+\frac{\mathrm{d}r^2}{f(r)}+r^2\mathrm{d}\Omega^2,
\end{equation}
where $\tau$ is the Euclidean time. To avoid a conical singularity
at the horizon, the Euclidean time  is made periodic with period
$1/T$, where $T$ is the Hawking temperature. Since we are interested
in a static solution with a radial scalar field, we can restrict
ourselves to a \textit{reduced} action. The latter can be obtained
by substituting the Euclidian metric~(\ref{euc}) in the
action~(\ref{eq:action}) and performing several integrations by
parts,
\begin{eqnarray}
\label{redaction} &&I_E=\int_{r_+}^\infty \text{d}r \left\{ -\frac{
N}{2 T}\Big[r(1-\beta e^{2\phi})+\left(2\alpha (3f-1)-\beta r^2
e^{2\phi}\right)\phi'+6f\alpha r(\phi')^2+2r^2\alpha
(\phi')^3f\Big]f'
\right.\nonumber\\
&&-\frac{N }{2 T}\Big[2f\left(2\alpha f-2\alpha-\beta r^2 e^{2\phi}\right)+8\alpha f^2 r\phi'+4r^2\alpha f^2(\phi')^2\Big]\phi''\\
&&-\frac{ N}{2 T}\left[-r^2\alpha f^2(\phi')^4+\left(2\alpha
f+2\alpha-\beta r^2 e^{2\phi}\right)f(\phi')^2-4\beta r f
e^{2\phi}\phi' -1+f+\beta e^{2\phi}(1-f)+\lambda r^2
e^{4\phi}\right] \left. \right\}+B\nonumber
\end{eqnarray}
Here, $B$ is a boundary term that is fixed by requiring that the solution of the equations of motion is an extremum of the Euclidean action. This condition implies that
\begin{eqnarray}
{\delta B}=\frac{N}{2 T}\Big[r(1-\beta e^{2\phi})+\left(2\alpha
(3f-1)-\beta r^2 e^{2\phi}\right)\phi'+ 6f\alpha
r(\phi')^2+2r^2\alpha (\phi')^3f\Big](\delta
f)+\left[\cdots\right](\delta \phi')+\left[\cdots\right](\delta\phi),
\end{eqnarray}
where the terms proportional to $(\delta\phi),(\delta \phi')$ are omitted for simplicity as they vanish identically on-shell. 
It is worth noticing an interesting feature of the solution we consider here. In general, the boundary term depends on the parameter $\beta$, as can be seen from the above equation. However, on-shell the terms proportional to $\delta\phi$ and $\delta\phi'$ drop out, while inside the first bracket, terms involving the $\beta$ parameter also cancel out. Therefore the resulting thermodynamic expression does not depend on $\beta$ for the solution (\ref{eq:solutionA})-(\ref{eq:solutionAbis}), as we will see below.
Indeed, on-shell the variation of the boundary term reduces to the following simple expression
\begin{eqnarray}
{\delta B}=\frac{1}{2 T r}\Big[2\alpha(1-f)+r^2\Big](\delta f).
\label{varB}
\end{eqnarray}
From the above expression it follows that
$$
(\delta B)|_{\infty}=-\frac{1}{2
 T}\left(1+\frac{\alpha}{r_+^2}\right)(\delta r_+)\quad \Longrightarrow \quad
B|_{\infty}= -\frac{1}{2 T}\left(r_+-\frac{\alpha}{r_+}\right)
$$
while for the variation at the horizon,
$$
(\delta B)|_{r_+}=-2\pi\left(r_++\frac{2\alpha}{r_+}\right)(\delta r_+) \quad \Longrightarrow \quad B|_{r_+}=-\pi\left[r_+^2+4\alpha \ln(r_+)\right].
$$
Hence, on-shell, the Euclidean action (\ref{redaction}) has value
\begin{eqnarray}
I_E=-\frac{1 }{2 T}\left(r_+-\frac{\alpha}{r_+}\right)+\left[\pi
r_+^2+4\pi\alpha\ln(r_+)\right]. \label{fineuc}
\end{eqnarray}
Comparing the above expression with the relation of the Euclidean
action to the mass ${\cal M}$ and the entropy ${\cal S}$ in the
grand canonical ensemble, $I_E=-\frac{\cal M} {T}+{\cal S}$, we find
that for the black hole
solution~(\ref{eq:solutionA})-(\ref{eq:solutionAbis}),
\begin{eqnarray}
{\cal M}=\frac{1 }{2}\left(r_+-\frac{\alpha}{r_+}\right)=M,\qquad
{\cal S}=\pi r_+^2+4\pi\alpha\ln(r_+). \label{parm}
\end{eqnarray}
Hence, one concludes that the usual one-quarter area law of the
entropy for general relativity is violated, as to be expected from
standard results in Lovelock gravity\cite{Myers:1988ze}{\footnote{In Lovelock gravity
the higher order term (in $\alpha$) provides a correction from the
induced curvature of the horizon surface while the GR term is simply
the tension associated to the horizon surface
\cite{Charmousis:2008kc}. This can be understood from the general
formalism of Iyer and Wald~\cite{Iyer:1994ys}.}}.
Nevertheless, the first law of thermodynamics holds, $d{\cal M}=Td{\cal S}$, with the Hawking
temperature given by
\begin{equation}
\label{temperature} T=\frac{r_+^2+\alpha}{4\pi r_+(2\alpha+r_+^2)}.
\end{equation}
As things stand we note that for $\alpha<0$, the temperature
diverges, i.e. $T\to \infty$ as $M$ goes to the minimal mass of the
black hole $ M_{\text{min}}$. This is not \textit{a priori} a
problem, however,  the free energy $F\equiv M-T\cal S$ then also diverges at a finite
mass. This can be remedied noting that the entropy is defined up to
a constant $s$, namely
\begin{equation}
{\cal S}_{\alpha<0} = \pi\left(r_+^2-2\left\lvert\alpha\right\rvert\ln\frac{r_+^2}{s\left\lvert\alpha\right\rvert}\right).
\end{equation}
We now fix $s = \frac{2}{\exp\left(1\right)}=\frac{2}{\rm e}$ to have vanishing entropy as $M\to M_{\text{min}}$ and therefore a
finite free energy (similar to the case of a Schwarzschild black hole in GR). 
For this choice of $s$, the free energy is positive (see also \cite{Clunan:2004tb}) and finite for any mass, decreasing from $M$ to $M/2$ as $M$ runs from $M_{\text{min}}$ to $\infty$. 

For positive $\alpha$ there is no lower limit on the black hole mass, and $T$ does not diverge for $M=0$. We can fix the free constant $s$ so that the entropy vanishes for the minimal mass $M= 0$, resulting in
\begin{equation}
{\cal S}_{\alpha>0} = \pi\left(r_+^2+2\alpha\ln\frac{r_+^2}{\rm{e}^{1/2}\alpha}\right).
\end{equation}
For $\alpha>0$ the free energy increases from $0$ to $M/2$ as $M$ runs from $0$ to $\infty$.
Let us  finally mention that, as for the Schwarzschild black hole, the heat capacity is negative for any sign of $\alpha$.

\subsection{A non trivial vacuum, the slowly rotating and Vaidya-like extensions}

As we pointed out in the beginning of the section, the
solution~(\ref{eq:solutionA})-(\ref{eq:solutionAbis}) does not reduce
to flat spacetime in the limit of zero black hole mass, $M\to 0$.
Moreover, as mentioned before, the zero mass spacetime has a
singularity at $r=0$ which is either naked ($\alpha<0$) or  covered
by a horizon ($\alpha>0$). One can also show that a trivial scalar field
does not lead to a flat spacetime solution. This means that any flat
geometric vacuum will necessarily require a non trivial
scalar field. Indeed, solving the field equations with a general
$\phi=\phi(t,r)$ and a flat metric, i.e. Eq.~(\ref{eq:solutionA}) with
$f=1$, we find two solutions, where the time-dependence of the scalar field must be non-trivial,
\begin{equation}
\phi(t) =
\ln\left(\frac{\sqrt{\left(-2\alpha/\beta\right)\left(3\pm\sqrt{6}\right)}}{\left\vert
t+\mu\right\vert}\right),\quad \phi(t,r)=
\ln\left(\frac{\sqrt{\left(-8\mu\alpha/\beta\right)\left(3\pm\sqrt{6}\right)}}{\left\vert
r^2-t^2+\mu\right\vert}\right),
\label{stealth}
\end{equation}
and  $\mu$ is an arbitrary constant. None of these profiles is differentiable in the whole spacetime. The solution~(\ref{eq:solutionA})-(\ref{eq:solutionAbis}) and the
flat configurations presented above cannot be smoothly deformed into
each other, which suggests that they belong to different,
disconnected sectors. Similar solutions have been discussed for non
minimally coupled scalar fields in Refs.~\cite{Ayon-Beato:2005yoq}
and \cite{Ayon-Beato:2004nzi}. In a somewhat different context, the
so-called Fab 4 theory, non-trivial flat vacua exist with
self-tuning properties \cite{Charmousis:2014mia}, although there is
no hint of self tuning within the presently considered theory.

It would be very interesting if one could generalize the static
solution~(\ref{eq:solutionA})-(\ref{eq:solutionAbis}) to its
stationary version. A fully analytic solution is not seemingly
easily found, one can however, as a first step, find the slowly
rotating solution in the manner described by Hartle and Thorne in GR
\cite{Hartle:1967he, Hartle:1968si}. The Hartle-Thorne
formalism in the presence of matter is very useful for calculating,
for example, the moment of inertia for neutron stars. In particular,
for most observed pulsars the Hartle-Thorne formalism is a good
approximation of their gravitational field. Here, in the absence of
matter, we will seek the slowly rotating version of our static
solution.

For the slowly rotating solution, we start with an ansatz for the metric of the form
\begin{equation}
\mathrm{d}s^2 = -f(r)\mathrm{d}t^2 +
\frac{\mathrm{d}r^2}{f(r)}+r^2\mathrm{d}\Omega^2-2\delta\,\omega(r)\,
r^2\sin^2\theta\mathrm{d}t\mathrm{d}\varphi, \label{eq:ansatzslow}
\end{equation}
where $\delta$ is a first order parameter, such that the angular
momentum per unit mass is given by $\delta a$ for slowly rotating
solutions. At first order, the only new contribution in the
equations of motion in comparison with the static case is the
off-diagonal $t\varphi-$component, while the geometric constraint
$R+\frac{\alpha}{2}\mathcal{G}=0$ is not affected at first order.
As a direct consequence, one finds that the metric function $f(r)$ and
the scalar field $\phi$ have the same profile (\ref{eq:solutionA})-(\ref{eq:solutionAbis}) as in the static case, while the solution for $\omega\left(r\right)$ is
\begin{equation}
\omega(r) = -6aM \int_\infty^r\frac{\mathrm{d}r}{r^4\sqrt{1+8\alpha\left(\frac{M}{r^3}+\frac{\alpha}{r^4}\right)}}.
\label{omega}
\end{equation}
As  $r\to\infty$, the integral gives to leading order the GR
behavior, with $\delta J=\delta aM$ the total angular momentum,
$$\omega\left(r\right)=\frac{2J}{r^3}\left[1-\frac{2\alpha
M}{r^3}-\frac{12 \alpha^2}{7 r^4}+
\mathcal{O}\left(\frac{1}{r^6}\right)\right],$$ and higher order correction
terms in $\alpha$.
 The variable $\omega$, as in GR, describes the speed at which a geodesic observer rotates because of frame dragging.

Yet another interesting feature of the static solution~(\ref{eq:solutionA})-(\ref{eq:solutionAbis}) within the
action~(\ref{eq:action}) is that it can be extended to a radiating
(or absorbing) Vaidya-like solution. The Vaidya solution in GR
describes a black hole with varying mass due to either radiation or
accretion of pressureless light-like matter. It is relevant, as a
paradigm for Hawking radiation or classically simulating
gravitational collapse of null dust. In the case of GR, the recipe
for the construction of the Vaidya solution is to use the retarded
$u$ (or advanced $v$) null coordinate, and then to promote the mass
parameter to a function of this null coordinate. In GR the Vaidya
solution contains a non-trivial energy-momentum tensor in the form
of light-like dust, whose only non-vanishing components are along
the retarded (or advanced) time. We will consider the same energy
momentum tensor here in addition to~(\ref{eq:action}). What turns
out to be crucial in finding the Vaidya extension is that the
trace of the effective energy-momentum tensor vanishes identically
(as so happens for an electromagnetic charge
\cite{Fernandes:2021dsb}). Therefore for our
action~(\ref{eq:action}), the geometric constraint
$R+\frac{\alpha}{2}\mathcal{G}=0$  is not modified in the presence
of minimally coupled  null dust.

Indeed, we find that the theory~(\ref{eq:action}) admits a radiating Vaidya extension,
\begin{equation}
\left\lbrace
\begin{split}
&\mathrm{d}s^2=-f\left(u,r\right)\mathrm{d}u^2-2\mathrm{d}u\mathrm{d}r+r^2\mathrm{d}\Omega^2,\quad f\left(u,r\right) =
1+\frac{r^2}{2\alpha}\left(1-\sqrt{1+8\alpha\left(\frac{M(u)}{r^3}+\frac{\alpha}{r^4}\right)}\right),\\
&\phi =\ln\left(\frac{\sqrt{-2\alpha/\beta}}{r}\right),\quad
T_{uu}=-\frac{M'(u)}{4\pi r^2}\geq 0,
\end{split}
\right. \label{eq:vaidyaA}
\end{equation}
as well as an accreting Vaidya extension,
\begin{equation}
\left\lbrace
\begin{split}
&\mathrm{d}s^2=-f\left(v,r\right)\mathrm{d}v^2+2\mathrm{d}v\mathrm{d}r+r^2\mathrm{d}\Omega^2,\quad f\left(v,r\right) =
1+\frac{r^2}{2\alpha}\left(1-\sqrt{1+8\alpha\left(\frac{M(v)}{r^3}+\frac{\alpha}{r^4}\right)}\right),\\
&\phi =\ln\left(\frac{\sqrt{-2\alpha/\beta}}{r}\right),\quad
T_{vv}=\frac{M'(v)}{4\pi r^2}\geq 0.
\end{split}
\right. \label{eq:vaidyaAin}
\end{equation}
The energy-momentum tensor, as in GR,
satisfies standard energy conditions. For example, the latter spacetime describes an accreting black hole that is
irradiated by null dust from mass $M_1$ to mass $M_2>M_1$. Here, for $\alpha<0$ we want $M_1>M_\text{min}$ in order for spacetime
to be well defined. As for GR, at each instant $v$ such that $M_1<M(v)<M_2$, the zeros of $f$ describe the location of the apparent horizon.
Note finally that whereas the radiating/accreting solutions of GR verify $R=0$, the solutions presented
here have non-zero scalar curvature and satisfy instead the relation $R+\frac{\alpha}{2}\mathcal{G}=0$.

\section{Extracting singularities by disformal transformation} \label{sec:WH}

Our findings in the previous section tell us that
solution~(\ref{eq:solutionA}) for $\alpha>0$ describes a black hole
with a singularity at $r=0$ always hidden by a horizon. In contrast,
for the choice $\alpha<0$, the solution always has a naked
singularity for sufficiently small masses $M<M_\text{min}=\frac{3
\sqrt{\left\vert\alpha\right\vert}}{2\sqrt{2}}$ and in particular for $M=0$. This may not
necessarily be a problem. Indeed it may be, that unlike GR, our
theory~(\ref{eq:action}) presents no mass gap between (neutron)
star solutions and black holes (see \cite{Charmousis:2021npl} for a
recent study where this mass gap is not present) or again, that
there exists another black hole solution with no such minimal mass
constraint. Either way, the existence of naked singularities is
surely an undesirable feature of a theory. In this section we will
consider two different ways of eliminating this problem using
disformal transformations. We will construct gravitating monopole-like and wormhole
solutions in beyond Horndeski theory, such that either spacetime is
regularized at the origin for $M=0$, or singularities for any $M$ are excised altogether from spacetime.

For the former case it was noted that ($M=0$) vacua, which were well behaved in
Horndeski theory, were developing singularities at the origin when
transformed via a disformal transformation in beyond Horndeski \cite{Bakopoulos:2022csr}. Here we
saw, quite the opposite for the initial (seed) solution in Horndeski theory i.e., that at the
origin our vacuum is ill-behaved as $f(0)\neq 1$. Can we fix the singularity present at the origin for $M=0$ by
disformal transformation to a beyond Horndeski theory?

For the latter case, wormholes were recently constructed in shift-symmetry Horndeski
theories with a throat that shrinks to zero as the mass parameter
goes to zero \cite{Bakopoulos:2021liw}. For the case of our interest we will seek solutions
that will have a well-defined and crucially {\it permanent} throat at $r=r_0$. Such an, {\it eternal} wormhole
will be shown to remove any naked singularity of the spacetime whatever the
mass parameter of the solution. Furthermore during this construction, we will
uncover a subtlety, concerning the action of the resulting beyond
Horndeski theory.

Let us consider disformal transformations of the following form,
\begin{equation}
\tilde{g}_{\mu\nu}=g_{\mu\nu}+D\left(\phi,X\right)\phi_\mu\phi_\nu,
\label{eq:disf}
\end{equation}
where $D$ is a function of both $\phi$ and of the kinetic term
$X=-\phi_\mu\phi^\mu/2$. If the disformal coefficient $D$ depends
only on $\phi$, $D=D(\phi)$, any Horndeski theory transforms into
another theory in the Horndeski class~\cite{Bettoni:2013diz}. On the
other hand, for more general transformations with $D=D(\phi,X)$, the
transformation~(\ref{eq:disf}) leads to extensions beyond Horndeski,
see \cite{Gleyzes:2014qga} and \cite{Zumalacarregui:2013pma}.
From an action point of view,
we can deduce that  one possible way to excise naked singularities
is to couple matter non-minimally to a particular disformed metric.
Or on the other hand, in terms of the new disformal metric to which
matter couples minimally, this amounts to making a disformal
transformation of the initial theory~(\ref{eq:action}) towards a new
(beyond Horndeski) theory.

For definiteness as our seed metric we consider a static black
hole~(\ref{eq:solutionA}-\ref{eq:solutionAbis}) with $\alpha<0$,
which for small enough mass has a naked singularity at $r_S=0$ or
$r_S=r_P$. Applying the disformal transformation~(\ref{eq:disf}) to
(\ref{eq:solutionA}), we find the disformed metric,
\begin{equation}
\mathrm{d}\tilde{s}^2=-f(r)\,\mathrm{d}t^2+\frac{\mathrm{d}r^2}{f(r)\,
W^{-1}\left(\phi,X\right)}+r^2\left(\mathrm{d}\theta^2+\sin^2\theta
\,\mathrm{d}\varphi^2\right),\label{eq:dstilde}
\end{equation}
where
$$
W\left(\phi,X\right)\equiv 1-2D\left(\phi,X\right)X.
$$
Note that, as usual, the resulting solution for the scalar $\phi$
remains unchanged and is given by~(\ref{eq:solutionAbis}).

\subsection{From a singular vacuum to a gravitational monopole-like solution}
As a first  working example, we will see that a  simple choice of the function $W\left(\phi,X\right)$ in (\ref{eq:dstilde})
enables to regularize the vacuum spacetime for $M=0$. 
Indeed, the metric solution (\ref{eq:solutionA}) admits
the following behavior at the origin
\begin{equation}
f\left(r\right)=1+\sqrt{2}-\frac{Mr}{\left\lvert\alpha\right\rvert\sqrt{2}}-
\left(1+\frac{M^2}{\left\lvert\alpha\right\rvert 2\sqrt{2}}\right)\frac{r^2}
{2\left\lvert\alpha\right\rvert}+\mathcal{O}\left(r^3\right),
\end{equation}
One can see that the vacuum metric $M=0$ would admit a
regular core if the value at the origin, $f(0)=1+\sqrt{2}$, could be rescaled to 1. A glance at the disformed metric
(\ref{eq:dstilde}) shows that choosing
$W\left(\phi,X\right)=1+\sqrt{2}$ enables to remove the pathologic
behaviour, yielding a disformal function $D\left(X\right) =-1/\left(\sqrt{2}X\right)$ and a new metric
\begin{equation}
\mathrm{d}\tilde{s}^2 = -\tilde{f}\left(r\right)\mathrm{d}t^2+\frac{\mathrm{d}r^2}
{\tilde{f}\left(r\right)}+r^2\mathrm{d}\Omega^2 \label{eq:mono}
\end{equation}
where $\tilde{f}\left(r\right) =
f\left(r\right)/\left(1+\sqrt{2}\right)$, and where the time coordinate
has been rescaled. Satisfyingly, this rescaling is not fine tuned, since it is independent of the theory parameter $\alpha$. The regularity of the resulting metric can be
better appreciated by looking at the Kretschmann scalar at $r=0$,
\begin{equation}
\tilde{R}_{\mu\nu\rho\sigma}\tilde{R}^{\mu\nu\rho\sigma}=\frac{4
\left(3-2 \sqrt{2}\right) M^2}{\alpha ^2 r^2}+ \frac{\left(6
\sqrt{2}-9\right) M \left(M^2-2 \sqrt{2} \alpha \right)}{\alpha ^3
r}+\mathcal{O}\left(1\right).
\end{equation}
Indeed, the diverging pieces of the Kretschmann invariant are now
proportional to $M$, boding well that the massless solution is now
regular. Of course, this naive rescaling of the metric at $r=0$ is
not without consequence on the nature of the spacetime
asymptotically: at  $r\to\infty$, the metric function
behaves as
\begin{equation}
\tilde{f}\left(r\right) = \sqrt{2}-1-\frac{2\left(\sqrt{2}-1\right)M}{r}+\mathcal{O}\left(\frac{1}{r^2}\right),
\end{equation}
such that at leading order, the asymptotic metric displays a solid angle deficit of $2\pi\left(1-\frac{1}{\sqrt{2}}\right)$, which is the characteristic signature of a global gravitating monopole \cite{Barriola:1989hx} embedded in GR. In summary, the metrics
(\ref{eq:mono}), parameterized by the integration constant $M$,
describes a regular, asymptotically monopole-like spacetime if $M=0$, a naked singularity in an asymptotically-monopolar background if $M<
M_{\text{min}}$, and a black hole in an asymptotically-monopolar background if $M\geq
M_{\text{min}}$. It is worth mentioning that the scalar field, which
is unchanged, diverges at $r=0$, although the spacetime is regular
in the massless case. A theory endowed with such scalar vacua would present very particular strong lensing properties, in particular double images \cite{Barriola:1989hx}. The associated beyond Horndeski theory is given in the appendix.

\subsection{An eternal wormhole excising a naked singularity}
We will now consider a general dependence of $D$ on both $\phi$ and $X$, and this will
be essential for the construction of wormhole solutions as well as the robust definition of the beyond Horndeski theory at hand.
To simplify expressions,  we redefine the scalar field as
\begin{eqnarray}
\psi=\sqrt{-\frac{2\alpha}{\beta}} \mathrm{e}^{-\phi}\Longrightarrow
\psi_{\mbox{\tiny{on-shell}}}=r, \label{defpsi}
\end{eqnarray}
with $\psi$ of  dimension 1.
We look for such $W\left(\psi,X\right)$ that the disformed metric (\ref{eq:dstilde}) describes a wormhole geometry. We have to impose three requirements on $W\left(\psi,X\right)$:
\begin{enumerate}
\item We require that $W^{-1}$ vanishes at a point $r=r_0$ such that $r_0>\{r_S,r_+\}$ if the spacetime admits a naked singularity $r=r_S$ or an event horizon $r=r_+$, so that $r=r_0$ corresponds to the wormhole throat, since
$\tilde{g}^{rr}\left(r_0\right)=0$ while
$\tilde{g}_{tt}\left(r\right)>0$ for any $r\geq r_0$.
\item The asymptotic flatness and the absence of solid deficit angle of the disformed metric is obtained by imposing that $W\to 1$ as $r$ goes to infinity.
\item The disformal transformation should be invertible, which implies that the determinant
of the Jacobian of the metric transformation~(\ref{eq:disf}) is not zero or infinity. This latter property is not
manifest in the solution itself but is essential for the robustness of the resulting beyond Horndeski action.
\end{enumerate}

\begin{figure}
\begin{subfigure}{0.51\textwidth}
\includegraphics[width=\linewidth]{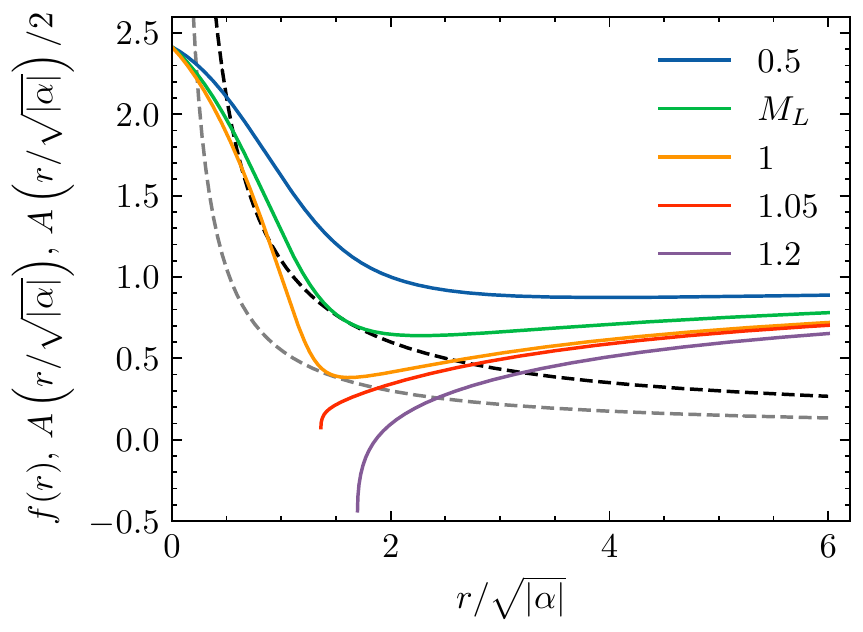}
\end{subfigure}
\begin{subfigure}{0.48\textwidth}
\includegraphics[width=\linewidth]{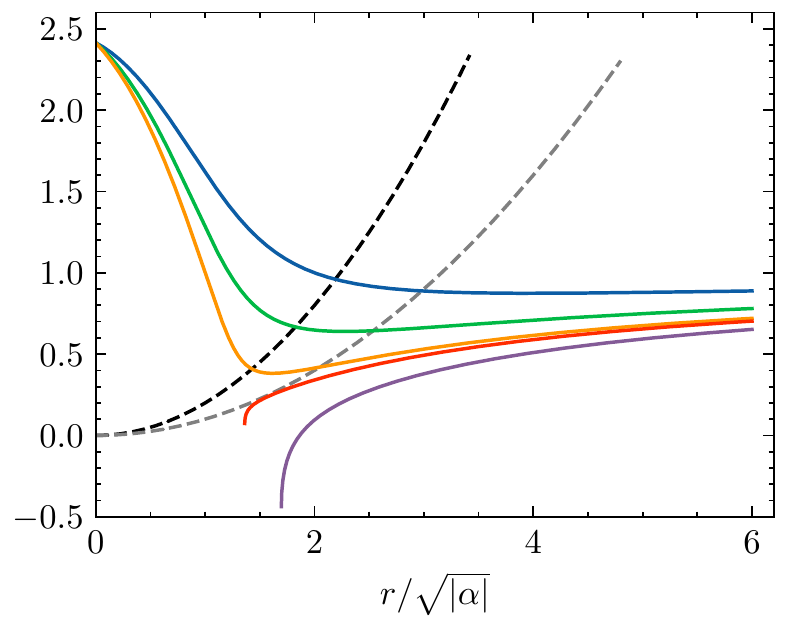}
\end{subfigure}
\caption{The functions $A$ (black curve) and $A/2$ (grey curve) are shown as functions of $r/\sqrt{|\alpha|}$ for two different cases: (\ref{eq:ourA}) with $a=0.1$ (left plot), and
$A\left(r/\sqrt{\left\lvert\alpha\right\rvert}\right)=r^2/\left(5\left\lvert\alpha\right\rvert\right)$ (right plot); while the metric function $f$ is shown for several values of $M/\sqrt{\left\lvert\alpha\right\rvert}$, in color. The throat radius $r_0$ (the singular radius $r_*$) is the largest intersection of $f$ with the black (grey) curve. On the left plot, $r_*$ is covered by the wormhole throat and the conditions for the disformal transformation formulated in the main text are satisfied. This is not the case for the right plot. The meaning of $M_L/\sqrt{\left\lvert\alpha\right\rvert}\approx 0.8213$ will be made clear later in the text.}
\label{fig:Aperfect}
\end{figure}

To this aim, we choose $W\left(\psi,X\right)$ to have the relatively simple form,
\begin{eqnarray}
W^{-1}\left(\psi,X\right)=\left(1-1/a\right)^{-1}\left(1+\frac{2\psi^2
X}{A\left(\psi/\sqrt{\left\lvert\alpha\right\rvert}\right)}\right).
\label{eq:condjacx}
\end{eqnarray}
The non-negative function $A\left(r/\sqrt{\left\lvert\alpha\right\rvert}\right)$ is such that $A\left(r\to\infty\right)=a$ where $a\neq 0,1$ in order for condition 2 to be fulfilled. 
Given that for our solution, $X=-\frac{f(r)}{2r^2}$, the throat $r=r_0$ of the wormhole is given at the
intersection of $f\left(r\right)$ with $A\left(r/\sqrt{\left\lvert\alpha\right\rvert}\right)$, namely
\begin{eqnarray}
f(r_0)=A\left(\frac{r_0}{\sqrt{\left\lvert\alpha\right\rvert}}\right). \label{eq:vanishing}
\end{eqnarray}
This is not all-the presence of the scalar field $\psi$, parameterized by the form of function $A$, is essential to guarantee that condition 3 is fullfilled as we will now see. 
Indeed condition 3 is not
manifest on the solution itself but is rather a requirement for the resulting
beyond Horndeski action. The disformal transformation becomes
non-invertible at two points. First at the throat $r=r_0$, due
to the infinite determinant of the transformed metric, the
disformed spacetime cannot be mapped to the original spacetime. This
is however a mere coordinate singularity as we will see below in
Eqs.~(\ref{goodcoords}) and (\ref{goodcoordsbis}). The second
singular point is given by the equation $1+2X^2 D_X=0$ where $D_X$
stands for the derivative with respect to $X$ of the disformal
factor (\ref{eq:disf}). For our choice of $W$ as in
(\ref{eq:condjacx}), this point is located at radius
$r=r_*$ such that
\begin{equation}
f\left(r_*\right) = \frac{1}{2} A\left(\frac{r_*}{\sqrt{\left\lvert\alpha\right\rvert}}\right). \label{eq:condjac}
\end{equation}
At $r=r_*$, the transformation~(\ref{eq:disf}) becomes
non-invertible since the determinant of the Jacobian becomes
infinite\footnote{As it is shown in the Appendix, the presence of
$r=r_*$ prevents the disformed metric from solving a well-defined
variational principle for the beyond Horndeski action, obtained via
the transformation~(\ref{eq:disf}).}, i.e. the condition 3 of the
above is not satisfied. In order for the wormhole solution to
originate from a unique well defined action, $A$ should be chosen
such that the location $r=r_*$ is smaller than the location $r=r_0$, that is $r_*<r_0$,
so that $r=r_*$ is also excised from the wormhole spacetime. This allows infinitely many possibilities for $A$, but for our purposes, one can easily prove that the simple choice
\begin{equation}
A\left(\frac{\psi}{\sqrt{\left\lvert\alpha\right\rvert}}\right) = a+\frac{\sqrt{\left\lvert\alpha\right\rvert}}{\psi} \label{eq:ourA}
\end{equation}
satisfies these requirements for any $0<a<1$. This is illustrated on the left plot of Fig.~\ref{fig:Aperfect}. Conversely, on the right plot, the disformal mapping $D$ does not depend on the scalar field, that is to say $A\propto\psi^2$ (see (\ref{eq:condjacx})). As a result condition 3 is not satisfied because the singularity of the disformal transformation at $r=r_*$ is hit before the throat, $r_0<r_*$. Note that the crossing point $r=r_*$ is not a singular point of the disformed metric, but the disformed metric ceases to solve well-defined field equations below $r=r_*$. 

At the end, the wormhole solution satisfying all three requirements reads (reinstating the original
scalar $\phi$),
\begin{eqnarray}
\mathrm{d}s^2 &=&-f\left(r\right)\mathrm{d}t^2+\frac{\mathrm{d}r^2}{h(r)}+r^2\mathrm{d}\Omega^2,\label{eq:wormholesolution}\\
\phi(r) &=&\ln\left(\frac{\sqrt{-2\alpha/\beta}}{r}\right),\label{eq:wormholesolution_bis}
\end{eqnarray}
where
\begin{equation}
\label{eq:wormholesolution_bisbis}
h(r)=\frac{f\left(r\right)}{1-1/a}\left(1-\frac{f\left(r\right)}{a+\frac{\sqrt{\left\lvert\alpha\right\rvert}}{r}}\right),
\end{equation}
and $f(r)$ is given in (\ref{eq:f}). The wormhole
configuration~(\ref{eq:wormholesolution}-\ref{eq:wormholesolution_bisbis})
is a solution of a beyond Horndeski theory (given in the appendix),
for any $M$. In addition to the parameters $\alpha$ and $\beta$ of
the original theory~(\ref{eq:action}), the new theory is also
parameterized by a dimensionless
parameter $a\in\left]0,1\right[$. 

One can compute the throat radius $r_0$ as a function of the mass $M$ of the wormhole, provided the function $A$ is invertible (which is of course the case for (\ref{eq:ourA})). Let $f_0$ be the value of the metric function at the throat, which essentially quantifies the
compactness of the wormhole,
\begin{equation}
f_0= f\left(r_0\right)=a +\frac{\sqrt{\left\lvert\alpha\right\rvert}}{r_0}. \label{eq:f0}
\end{equation}
Indeed, if $f_0\ll 1$, then\footnote{We will see that $r_0\to \infty$ for large $M$, so $f_0\sim a$, and $f_0\ll 1$ happens if $a\ll 1$.} the redshift is important and the wormhole behaves very much like a black hole horizon for far away observers (see for example \cite{Damour:2007ap}). Equation (\ref{eq:f0}) enables us to get $r_0$ and $M$ as functions of $f_0$. Cautiously inverting the latter relation yields $f_0$ as a function of $M$, which finally gives $r_0$ as a function of $M$. 
 
\begin{figure}
\begin{subfigure}{0.51\textwidth}
\includegraphics[width=\linewidth]{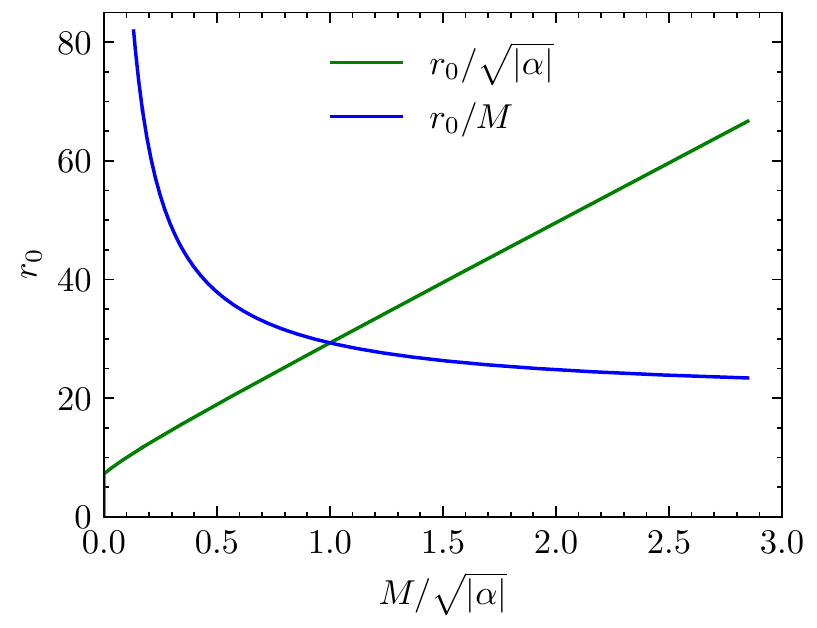}
\end{subfigure}
\begin{subfigure}{0.48\textwidth}
\includegraphics[width=\linewidth]{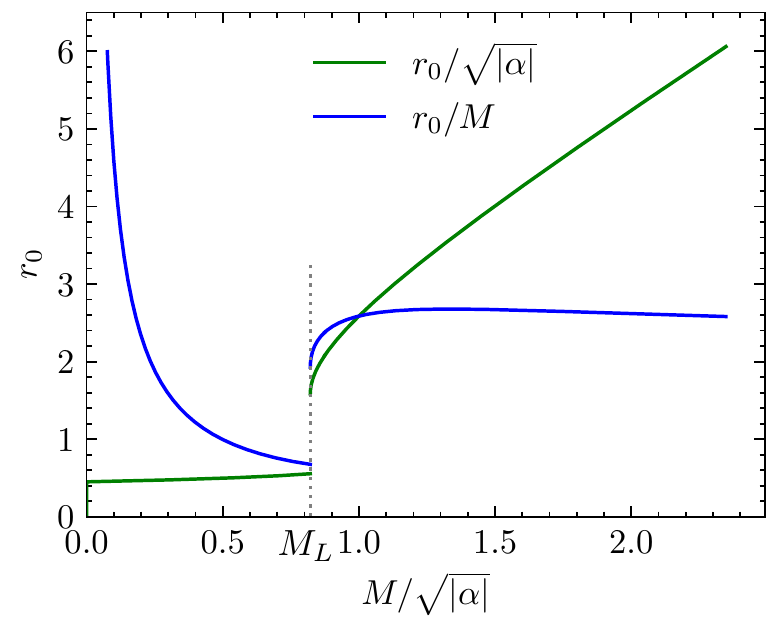}
\end{subfigure}
\caption{The plot shows the throat radius $r_0$ as a function of
$M/\sqrt{|\alpha|}$, for $a=0.9$ (left plot, no discontinuity in $r_0$) and $a=0.1$ (right plot, discontinuity at $M_L$). The discontinuity corresponds to a change of branch in
the solution of (\ref{eq:vanishing}). In~Fig.~\ref{fig:Aperfect}, different
branches correspond to intersections of $A$ (black dashed curve) and
$f$ (colored curves).}
\label{fig:rzero}
\end{figure}

This procedure enables to show that there exists a value\footnote{More precisely, $a_0$ is the unique root in $\left]0,1\right[$ of the equation $-1127+2956a-2948a^2+1532a^3-120a^4-480a^5+224a^6-32a^7=0$.} $a_0\approx 0.87396$ of the parameter $a$, such that for $a\geq a_0$, $r_0$ is a smooth function of $M$, while for $a<a_0$, there is a discontinuity in $r_0$ at a mass $M_L$ (which depends of course on $a$). Fig.~\ref{fig:rzero} illustrates these different behaviours for the values $a=0.9$ (left plot) and $a=0.1$ (right plot). One can easily understand this behaviour by taking a look at the left plot of Fig.~\ref{fig:Aperfect}, which corresponds to $a=0.1$: for $M<M_L$ (blue curve), the throat is close to the origin and blueshifted, while for $M>M_L$ (yellow curve), the throat is at a bigger radius and redshifted. 

Obviously, the size of the throat increases with the parameter $a$. For example, it is easy to show that the throat radius quickly converges towards $r_0\approx 2M/\left(1-a\right)$ as soon as $M>\sqrt{\left\lvert\alpha\right\rvert}$ (which corresponds at most to the order of magnitude $M>10M_\odot$, according to the bounds on $\left\lvert\alpha\right\rvert$ given in the previous section). Hence a throat radius enhanced by a factor $\left(1-a\right)^{-1}$ with respect to the Schwarzschild radius for the corresponding mass.
\\

\begin{figure}
\begin{subfigure}{0.495\textwidth}
\includegraphics[width=\linewidth]{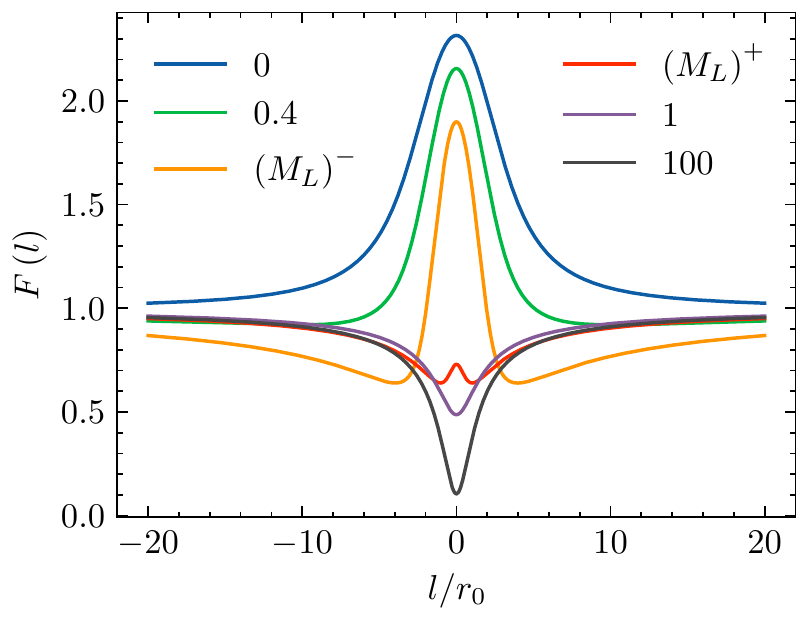}
\end{subfigure}
\begin{subfigure}{0.495\textwidth}
\includegraphics[width=\linewidth]{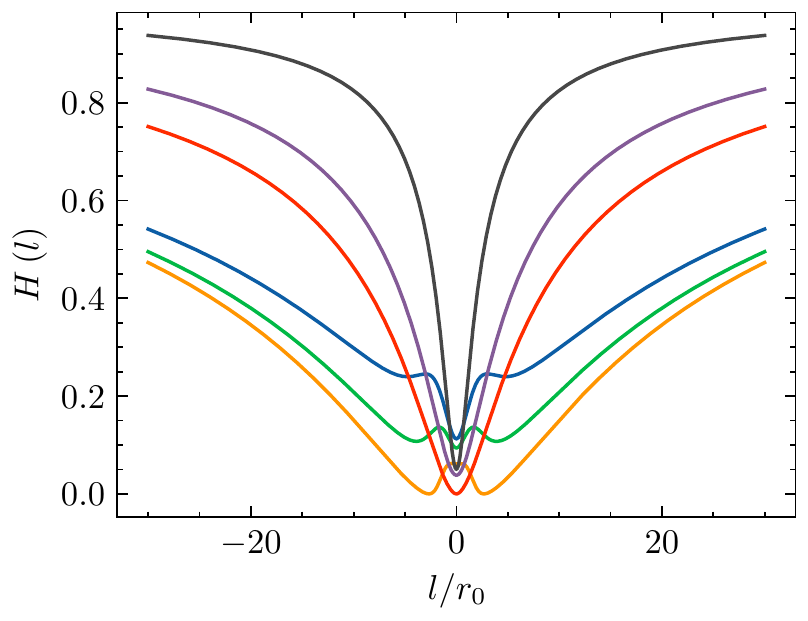}
\end{subfigure}
\caption{Functions $F\left(l\right)$ and $H\left(l\right)$ of
metric (\ref{goodcoords}) (with parameter $a=0.1$), for different values of
$M/\sqrt{\left\lvert\alpha\right\rvert}$ given by the legend. The values $\left(M_L\right)^-$ and $\left(M_L\right)^+$ are as close as possible to the limit mass $M_L$ with our numerical precision, namely $\left(M_L\right)^\pm=M_L\left(1\pm 10^{-15}\right)$, illustrating the discontinuity occurring at this mass. For huge masses, the redshift function converges to the value $a$($=0.1$ here) at the throat.}\label{fig:f2f3}
\end{figure}

We conclude our discussion by presenting the wormhole solutions using everywhere non-singular coordinates (including the throat).
To do this we change the radial coordinate $r$ by introducing $l$ with range $l\in\left]-\infty,\infty\right[$ defined by
\begin{equation}
r^2 = l^2 + r_0^2
\end{equation}
In this coordinate system, any wormhole metric, with throat $r_0$ of the form (\ref{eq:wormholesolution}), is given by
\begin{equation}
\label{goodcoords}
\mathrm{d}s^2 = -F\left(l\right)\mathrm{d}t^2+\frac{\mathrm{d}l^2}{H\left(l\right)}+\left(l^2+r_0^2\right)\mathrm{d}\Omega^2,
\end{equation}
where
\begin{equation}
\label{goodcoordsbis}
F\left(l\right) = f\left(\sqrt{l^2+r_0^2}\right),\quad
H\left(l\right) = h\left(\sqrt{l^2+r_0^2}\right)\frac{l^2+r_0^2}{l^2}.
\end{equation}
Note that the function $H\left(l\right)$ is regular everywhere, and in particular at the throat $l\to 0$ we have,
\begin{equation}
H\left(l\right)=\frac{r_0}{2}h'\left(r_0\right)+\mathcal{O}\left(l^2\right).
\label{eq:f3lf1prime}
\end{equation}
Since $h\left(r>r_0\right)> 0$, hence $H\left(l\right)\geq0$
everywhere\footnote{$H\left(l\right)= 0$ occurs for $l=0$ and $h'\left(r_0\right)=0$. This corresponds to the particular value of $M$ where a discontinuity in $r_0$ occurs, see  Fig.~\ref{fig:rzero}.}.
The other metric function, $F\left(l\right)$, is regular and non-negative everywhere. In Fig.~\ref{fig:f2f3}, we plot the
functions $F\left(l\right)$ and $H\left(l\right)$ for different
masses $M$, when $a=0.1$. The masses of the yellow and red plots are chosen very
close to the mass $M_L$ where occurs the $r_0$ discontinuity: for
the yellow plot, the mass is still sufficiently low so that the
throat $r_0$ is close to $r=0$ and blueshifted,
while for the red plot, the throat $r_0$ is much larger and the
spacetime is redshifted there. This is not just a sharp evolution of
the $F\left(l\right)$ behavior as a function of the mass, but a
true discontinuity at $M=M_L$. 

\section{Conclusions}
In this paper we have  studied solutions of the theory~(\ref{eq:action}) as well as certain of its disformal versions.
The theory~(\ref{eq:action}) is in the class of Horndeski theory, and, thanks to underlying symmetries as well as a particular 
choice of relation between coupling constants, exact solutions can be found analytically.

We analyzed in detail the metric of a spherically symmetric
solution~(\ref{eq:solutionA}-\ref{eq:solutionAbis}), first found
in~\cite{Fernandes:2021dsb}. Depending on the sign of the coupling
$\alpha$ (and hence $\lambda$), the physical meaning of the solution
may differ drastically. For positive $\alpha$ the spacetime
(\ref{eq:solutionA}) with (\ref{eq:f})  always describes a black
hole with a singularity hidden by a horizon, similar to GR black
holes. It is worth noting that for $\alpha>0$, either other spherically symmetric solutions describing spacetime outside a gravitating body exist, either $\alpha$ satisfies the tight constraint~(\ref{eq:constraint}), implying virtually no modifications
of GR for any present-day and near future observations. The case of
$\alpha<0$ is more involved. Indeed, in this case there is a
limiting mass $M_\text{min}$ given in terms of the parameters of the
theory, Eq.~(\ref{M0}). For  $M>M_\text{min}$, the
spacetime~(\ref{eq:solutionA}) with (\ref{eq:f}) describes a black
hole. For $M\leq M_\text{min}$, the
solution~(\ref{eq:solutionA}), (\ref{eq:f}) corresponds to a naked
singularity.

The analysis of the  black hole thermodynamics showed that the
entropy of the black hole receives a $\log$-correction,
Eq.~(\ref{parm}), that depends only on the parameter $\alpha$ of the
theory. Meanwhile, the first law of thermodynamics holds, with the
Hawking temperature given by~(\ref{temperature}), that also depends
on the coupling $\alpha$, while the mass is indeed given by $M$.

We then presented three new classes of solutions of~(\ref{eq:action}). The first type is a non trivial flat solution,
given by Eq.~(\ref{stealth}). The solution has a non-trivial scalar
field, while the metric remains flat, i.e. the backreaction of the
scalar field is absent in this case. The second solution is an
extension of the black hole solution~(\ref{eq:solutionA}),
(\ref{eq:f})  to a slowly rotating case, Eqs.~(\ref{eq:ansatzslow}),
(\ref{omega}). Probably the most interesting case is the third new
solution we found, an analogue of the Vaidya solution of GR. The
solutions~(\ref{eq:vaidyaA}) and (\ref{eq:vaidyaAin}) describe
correspondingly radiating and accreting solutions of the
theory~(\ref{eq:action}), that are counterparts of the Vaidya
solution in GR. The mass of the black hole $M=M(v)$ ($M=M(u)$) grows
(decreases) due to the infall (radiation) of light dust.

The last part of the paper is devoted to the disformal
transformations of theory~(\ref{eq:action}) and its solutions. We
focused on the case $\alpha<0$ where the theory admits naked
singularities for small enough masses $M<M_\text{min}$. We proposed a remedy to avoid
the pathology by coupling matter to a disformed metric, which amounts
to making a disformal transformation of the theory~(\ref{eq:disf}). We first showed that a very simple choice of disformal parameter $D=D(X)$ led to a theory admitting gravitating monopole-like solutions, and where the $M=0$ spacetime is regular at $r=0$. On the other hand, we found a general form of the disformal parameter $D=D(\phi,X)$,
such that the naked singularity of the original theory is
transformed to a wormhole whose metric is regular everywhere, for any mass $M$. An
interesting feature of the obtained solutions is that wormholes
with both redshift and blueshift at the throat exist. The blueshift
at the throat implies that the light traveling through a wormhole
experiences blueshift as it approaches the throat, which is in
contrast to the standard behaviour, e.g. in the case of GR, when
light is always redshifted near gravitating sources.

Several questions arise on other choices of disforming functions $D(\phi,X)$, as well as 
the analysis of stability for the obtained wormhole solutions. It has been shown before that there are no stable wormholes in Horndeski theory~\cite{Evseev:2017jek}, 
while the extensions of Horndeski theory have a chance to support stable wormholes~\cite{Mironov:2018uou,Franciolini:2018aad}. Therefore it remains to be seen whether our wormhole solutions in beyond Horndeski theory are stable or not.
It would be also important to explore in detail observational features of the wormholes, such as light rings, shadows, and contrast them with compact objects of GR. It would also be interesting to look for stationary metrics within this theory~(\ref{eq:action}). The presence of an always valid geometric constraint may give a hint on the form of stationary solutions. Last but not least it would be interesting to study neighbouring theories to~(\ref{eq:action}) and find spherically symmetric solutions there. These are some of the intriguing questions we hope will be studied in the near future. 

\acknowledgements{We are very happy to thank Timothy Anson and Karim Noui  for useful discussions, as well as Athanasios Bakopoulos and Panagiota Kanti for their insightful remarks regarding construction of wormholes. The authors also gratefully acknowledge the kind support of the PROGRAMA DE COOPERACI\'ON CIENT\'IFICA ECOSud-CONICYT 180011/C18U04. The work of MH has been partially supported by FONDECYT grant 1210889.}

\appendix

\section{Theories and solutions arising from the initial action }
\subsection{Known solutions}
We evoked in the introduction the existence of other relevant theories arising from the original action (\ref{eq:action}), with $\lambda=3\beta^2/\left(4\alpha\right)$ or $\beta=0=\lambda$. 
It was shown in~\cite{Fernandes:2021dsb} (see also~\cite{Charmousis:2021npl}) that they admit the following asymptotically flat, spherically symmetric solution:
\begin{equation}
\mathrm{d}s^2=-f(r)\mathrm{d}t^2+\frac{\mathrm{d}r^2}{f(r)}+r^2\mathrm{d}\Omega^2,\quad f(r) = 1+\frac{r^2}{2\alpha}\left(1-\sqrt{1+\frac{8\alpha M}{r^3}}\right), \label{eq:solutionB}
\end{equation}
for any ADM mass $M$, along with the respective scalar field profiles:
\begin{equation}
\phi = \ln\left(\frac{\sqrt{-2\alpha/\beta}}{r}\right)-\ln \cosh \left(c_3 \pm \int \frac{\mathrm{d}r}{r\sqrt{f}}\right),\quad \phi=\int\mathrm{d}r\frac{\pm 1-\sqrt{f}}{r\sqrt{f}}. \label{eq:twoscalarfield}
\end{equation}
The scalar field constant $c_3$ is unconstrained, while the second profile is defined up to an additive constant, since (\ref{eq:action}) with $\beta=0=\lambda$ is the shift-symmetric four dimensional Einstein-Gauss-Bonnet (EGB) theory, see \cite{Fernandes:2020nbq}. We can thus, for this latter theory, add a linear time dependence for the scalar field: $\phi = \mu t + \psi(r)$, with $\mu$ a constant, without breaking the spherical symmetry of the scalar field derivatives. This was done in~\cite{Charmousis:2021npl} and leads to
\begin{equation}
\psi = \int\mathrm{d}r\frac{\pm\sqrt{\mu^2r^2+f}-f}{rf}, \label{eq:egbtime}
\end{equation}
and one finds that for any $\mu$, this profile is solution, along with an unchanged spacetime (\ref{eq:solutionB}). For $\mu=0$, the linear time dependence disappears, and one recovers the previous profile of (\ref{eq:twoscalarfield}). 
\\
We will now, in a similar fashion to the body of the paper, focus on flat spacetime, slowly rotating and radiating solutions for the above two theories. 

\subsection{Flat spacetime solutions}

As opposed to what we studied in the main text, the obtained spacetime (\ref{eq:solutionB}) does reduce to flat spacetime as $M\to 0$, that is to say $f(r)\to 1$. In this case, the scalar fields of (\ref{eq:twoscalarfield}) reduce to:
\begin{equation}
\phi = \ln\left(\frac{\mu\sqrt{-8\alpha/\beta}}{1+\mu^2r^2}\right)
\end{equation}
where $\mu=\exp\left(\pm c_3\right)$ for the first one, and:
\begin{equation}
\phi = 0 \quad\text{or}\quad \phi = -2\ln r 
\end{equation}
up to an additive constant for the second one, for the respective choice of plus or minus sign. As regards the solution (\ref{eq:egbtime}) with $\phi = \mu t+\psi(r)$, it corresponds to the same spacetime and therefore gives another possibility for a stealth flat spacetime solution as $M\to 0$, with a scalar field reducing to:
\begin{equation}
\phi = \mu t -\ln r \pm\left(\sqrt{\mu^2r^2+1}-\arctanh\sqrt{\mu^2r^2+1}\right).
\end{equation} 
We can nevertheless question if other flat spacetime solutions, with $\phi=\phi\left(t,r\right)$, exist. We find the following solutions: on the one hand, when $\lambda=3\beta^2/\left(4\alpha\right)$,
\begin{align}
\phi ={}& \phi(r) = \ln\left(\frac{\mu\sqrt{-8\alpha/\beta}}{1+\mu^2r^2}\right),\\
\phi ={}& \phi(t) = \ln\left(\frac{\sqrt{-2\alpha/\beta}}{\left\lvert t+\mu\right\rvert}\right),\\
\phi ={}& \phi(t,r) = \ln\left(\frac{\sqrt{-8\mu\alpha/\beta}}{\left\lvert r^2-t^2+\mu\right\rvert}\right).
\end{align}
The first line, as shown above, comes directly from the black hole scalar field as $M\to 0$, while the other lines are different branches. In each case, $\mu$ is an integration constant. Only the first branch is differentiable in the whole spacetime. On the other hand, when $\beta=0=\lambda$, one gets up to a constant,
\begin{align}
\phi = {}& 0,\\
\phi ={}& \phi(r) = -2\ln r \label{eq:psilnr},\\
\phi ={}& \phi(t,r) = \mu t -\ln r \pm\left(\sqrt{\mu^2r^2+1}-\arctanh\sqrt{\mu^2r^2+1}\right)\label{eq:psimutflat},\\
\phi ={}& \phi(t,r) = -\ln\left\lvert r^2-t^2\right\rvert.
\end{align}
The only new solution not described above is the last one. The constant profile and the $+$ branch of (\ref{eq:psimutflat}) are differentiable for any $r\geq 0$.

\subsection{Slowly rotating solutions}
Let's now turn to the slowly rotating solutions. The ansatz metric is the same (\ref{eq:ansatzslow}) as in the main text, and the same discussion is still valid: one gets the same $f(r)$ (\ref{eq:solutionB}) and scalar fields (\ref{eq:twoscalarfield}) (or also the time-dependent scalar field $\phi =\mu t+\psi\left(r\right)$, (\ref{eq:egbtime})) as in spherical symmetry. Finally, $\omega\left(r\right)$ is given by
\begin{equation}
\omega(r) =  -6aM \int_\infty^r\frac{\mathrm{d}r}{r^4\sqrt{1+\frac{8\alpha M}{r^3}}} =-\frac{a}{2\alpha}\left(1-\sqrt{1+\frac{8\alpha M}{r^3}}\right), \label{eq:omega2sol}
\end{equation}
where, once again, the GR limit is fulfilled asymptotically. The slowly rotating metric is therefore the same for both theories, with different scalar fields. Note that, for $\beta=0=\lambda$, the slowly rotating solution has already been given in \cite{Charmousis:2021npl}.

\subsection{Radiating solutions}
We proceed with the Vaidya-like solutions. While we ended up with an unchanged spherically-symmetric scalar field in the body of the paper, this is no longer the case: the dependence of the scalar field on the null coordinate $u$ or $v$ is no longer trivial. In fact, one finds that the scalar field must satisfy a non-linear partial differential equation (PDE) which does not admit any obvious solution. But, assuming this PDE is satisfied, i.e. taking it as an implicit definition for the scalar field, all field equations are satisfied, and one ends up with the following outgoing-Vaidya-like solution
\begin{equation}
\left\lbrace
\begin{split}
&\mathrm{d}s^2=-f\left(u,r\right)\mathrm{d}u^2-2\mathrm{d}u\mathrm{d}r+r^2\mathrm{d}\Omega^2,\quad f\left(u,r\right) = 1+\frac{r^2}{2\alpha}\left(1-\sqrt{1+\frac{8\alpha M(u)}{r^3}}\right),\\
&0=2 \alpha  \left(f \left(r \phi'+1\right)^2-2 r \dot{\phi} \left(r \phi'+1\right)-1\right)-\beta  r^2 e^{2 \phi },\quad T_{uu}=-\frac{M'(u)}{4\pi r^2}\geq 0,
\end{split}
\right. \label{eq:vaidyaB}
\end{equation} 
and the following ingoing-Vaidya-like solution
\begin{equation}
\left\lbrace
\begin{split}
&\mathrm{d}s^2=-f\left(v,r\right)\mathrm{d}v^2+2\mathrm{d}v\mathrm{d}r+r^2\mathrm{d}\Omega^2,\quad f\left(v,r\right) = 1+\frac{r^2}{2\alpha}\left(1-\sqrt{1+\frac{8\alpha M(v)}{r^3}}\right),\\
&0=2 \alpha  \left(f \left(r \phi'+1\right)^2+2 r \dot{\phi} \left(r \phi'+1\right)-1\right)-\beta  r^2 e^{2 \phi },\quad 
T_{vv}=\frac{M'(v)}{4\pi r^2}\geq 0.
\end{split}
\right. \label{eq:vaidyaBin}
\end{equation} 
The PDE taken as an implicit definition of the scalar field is given below the metric, and with, of course, $\beta=0$ for the shift-symmetric four dimensional EGB case. A prime denotes derivation with respect to $r$, while a dot stands for derivation with respect to $u$ or $v$.

\section{Disformal transformations}

The gravitating monopole-like solution solves the following beyond
Horndeski theory, where for readability, the variables $\phi$ and
$\tilde{X}$ (the disformed kinetic term) are replaced respectively
by $y$ and $x$,
\begin{align*}
\tilde{G_2}\left(y,x\right) ={}& 8 \sqrt{5 \sqrt{2}+7} \alpha  x^2+12 \sqrt{\sqrt{2}+1} \beta  x \mathrm{e}^{2 y}-\frac{\sqrt{\sqrt{2}-1}
\beta ^2 \mathrm{e}^{4 y}}{2 \alpha }+8 \sqrt{2 \left(\sqrt{2}-1\right)} \beta  x \mathrm{e}^{2 y} \ln \left\lvert x\right\rvert,\\
\tilde{G_3}\left(y,x\right) ={}& 8 \left(\sqrt{2}+1\right)^{3/2} \alpha  x+4 \sqrt{2 \left(\sqrt{2}-1\right)} \beta  \mathrm{e}^{2 y}+2
\sqrt{2 \left(\sqrt{2}-1\right)} \beta  \mathrm{e}^{2 y} \ln \left\lvert x\right\rvert,\\
\tilde{G_4}\left(y,x\right) ={}& \sqrt{\sqrt{2}-1}+4 \sqrt{\sqrt{2}+1} \alpha  x-\sqrt{\sqrt{2}-1} \beta  \mathrm{e}^{2 y},\\
\tilde{G_5}\left(y,x\right) ={}& \frac{4 \alpha  \ln \left\lvert x\right\rvert}{\sqrt{\sqrt{2}-1}},\\
\tilde{F_4}\left(y,x\right) ={}& \frac{\sqrt{\sqrt{2}-1} (\beta  \mathrm{e}^{2 y}-1)}{2 \sqrt{2} x^2}+\frac{\sqrt{2 \left(\sqrt{2}+1\right)} \alpha }{x},\\
\tilde{F_5}\left(y,x\right) ={}& \frac{\left(\sqrt{2}-2\right) \alpha }{3 \left(\sqrt{2}-1\right)^{3/2} x^2}.
\end{align*}
The main differences (apart from the beyond Horndeski terms) with
the original theory (\ref{couplingfcts}) are the terms proportional
to $\ln\vert x\vert$ in $\tilde{G_2}$ and $\tilde{G_3}$.

More generally, we now present the disformed Horndeski action
which arises through a disformal transformation (\ref{eq:disf}) of a
general starting Horndeski action (\ref{eqaction}-\ref{eqaction1}).
The disformed Horndeski action belongs to the so-called beyond
Horndeski theory and is given by
\begin{eqnarray}
S = \int \mathrm{d}^4x\sqrt{-\tilde{g}}\left\lbrace
\mathcal{\tilde{L}}_2+\mathcal{\tilde{L}}_3+\mathcal{\tilde{L}}_4+\mathcal{\tilde{L}}_5+\mathcal{\tilde{L}}_{4b}+\mathcal{\tilde{L}}_{5b}\right\rbrace,
\end{eqnarray}
where appear the two additional beyond Horndeski Lagrangians that
read
\begin{eqnarray*}
&&\mathcal{\tilde{L}}_{4b} = \tilde{F_4}
\left(\phi,\tilde{X}\right)\left\lbrace 2
\tilde{X}\left[\left(\tilde{\Box\phi}\right)^2
-\left(\tilde{\phi}_{\mu\nu}\right)^2 \right]+
2\left[\tilde{\Box\phi}\tilde{\phi}^\mu\tilde{\phi}_{\mu\nu}\tilde{\phi}^\nu-\tilde{\phi}_\mu\tilde{\phi}^{\mu\nu}\tilde{\phi}_{\nu\rho}\tilde{\phi}^\rho\right]
\right\rbrace,   \\
&&\mathcal{\tilde{L}}_{5b}
=\tilde{F_5}\left(\phi,\tilde{X}\right)\left\lbrace
2\tilde{X}\left[\left(\tilde{\Box\phi}\right)^3 - 3\tilde{\Box\phi}
\left(\tilde{\phi}_{\mu\nu}\right)^2 +2\tilde{\phi}_{\mu\nu}\tilde{\phi}^{\nu\rho}\tilde{\phi}{^\mu_\rho}\right]\right.\\
&&\left.\qquad+3\left[\left(\tilde{\Box\phi}\right)^2\tilde{\phi}^\mu\tilde{\phi}_{\mu\nu}\tilde{\phi}^\nu-2\tilde{\Box\phi}
\tilde{\phi}_\mu\tilde{\phi}^{\mu\nu}\tilde{\phi}_{\nu\rho}\tilde{\phi}^\rho-\tilde{\phi}_{\mu\nu}\tilde{\phi}^{\mu\nu}
\tilde{\phi}^\rho\tilde{\phi}_{\rho\sigma}\tilde{\phi}^\sigma+2\tilde{\phi}_\mu\tilde{\phi}^{\mu\rho}\tilde{\phi}_{\rho\nu}
\tilde{\phi}^{\nu\sigma}\tilde{\phi}_\sigma\right]\right\rbrace,
\end{eqnarray*}
where $\tilde{\phi}_\mu=\tilde{\nabla}_\mu\phi$ and
$\tilde{X}=\frac{X}{1-2DX}$. The disformed Horndeski functions
$\tilde{G_k}\left(\phi,\tilde{X}\right)$ are given by
\begin{align*}
\tilde{G_2}={}&G_2\sqrt{1+2D\tilde{X}}-2\tilde{X}\left(H_3+H_4+H_5\right)_\phi-\frac{2\tilde{X}^2G_3D_\phi}{\left(1+2D\tilde{X}\right)^{3/2}},\\
\tilde{G_3}={}&\frac{G_3}{\sqrt{1+2D\tilde{X}}}-\left(H_3+H_4+H_5\right)\\
{}&\quad+2\tilde{X}\left\lbrace H_{R,\phi\phi}-H_{\Box,\phi}+\frac{1}{\sqrt{1+2D\tilde{X}}}
\left[2DG_{4\phi}-D_\phi\left(\frac{2\tilde{X}G_{4\tilde{X}}}{1-2\tilde{X}^2D_{\tilde{X}}}-G_4\right)\right]\right\rbrace,\\
\tilde{G_4}={}&G_4\sqrt{1+2D\tilde{X}}+\tilde{X}\left(H_{R,\phi}-\frac{\tilde{X}G_5D_\phi}{\left(1+2D\tilde{X}\right)^{3/2}}\right),\\
\tilde{G_5}={}&\frac{G_5}{\sqrt{1+2D\tilde{X}}}+H_R,
\end{align*}
while the beyond Horndeski functions
$\tilde{F_k}\left(\phi,\tilde{X}\right)$ read
\begin{align*}
\tilde{F_4}={}&\frac{D_{\tilde{X}}}{2}\left(\frac{2\tilde{X}G_{4\tilde{X}}\sqrt{1+2D\tilde{X}}}
{1-2\tilde{X}^2D_{\tilde{X}}}-\frac{G_4}{\sqrt{1+2D\tilde{X}}}\right)-\frac{1}{2}H_{R,\phi\tilde{X}}-
\frac{\tilde{X}^3G_{5\tilde{X}}D_{\tilde{X}}D_\phi}{\left(1-2\tilde{X}^2D_{\tilde{X}}\right)\left(1+2D\tilde{X}\right)^{3/2}},\\
{}&\quad+\frac{G_{5\phi}D}{2\left(1+2D\tilde{X}\right)^{3/2}}+\frac{G_5}{2\left(1+2D\tilde{X}\right)^{5/2}}\left\lbrace\tilde{X}
\left(1+2D\tilde{X}\right)D_{\phi\tilde{X}}+D_\phi\left[1-\tilde{X}\left(D+3\tilde{X}D_{\tilde{X}}\right)\right]\right\rbrace\\
\tilde{F_5}={}&-\frac{\tilde{X}G_{5\tilde{X}}D_{\tilde{X}}}{6\left(1-2\tilde{X}^2D_{\tilde{X}}\right)\sqrt{1+2D\tilde{X}}}.
\end{align*}
For clarity, we  have defined the following functions
\begin{equation*}
H_\Box =
\frac{\tilde{X}G_5D_\phi}{\left(1+2D\tilde{X}\right)^{3/2}},\quad
H_R =
\int\mathrm{d}\tilde{X}\frac{G_5\left(D+\tilde{X}D_{\tilde{X}}\right)}{\left(1+2D\tilde{X}\right)^{3/2}},\quad
H_5=\int\mathrm{d}\tilde{X}\left(H_{\Box,\phi}-H_{R,\phi\phi}\right),
\end{equation*}
and
\begin{equation*}
H_3 =
\int\mathrm{d}\tilde{X}\frac{-G_3\left(D+\tilde{X}D_{\tilde{X}}\right)}{\left(1+2D\tilde{X}\right)^{3/2}},\quad
H_4=\int\frac{\mathrm{d}\tilde{X}}{\sqrt{1+2D\tilde{X}}}\left[D_\phi\left(\frac{2\tilde{X}G_{4\tilde{X}}}
{1-2\tilde{X}^2D_{\tilde{X}}}-G_4\right)-2DG_{4\phi}\right],
\end{equation*}
thus following the notations of \cite{Bettoni:2013diz}, with the difference that we are including an $X$ dependence for the disformal function.

Let us now apply this disformal transformation to our specific
action (\ref{eq:action}) and its solution
(\ref{eq:solutionA}-\ref{eq:solutionAbis}) with the following choice
of $W^{-1}$,
\begin{equation}
W^{-1}\left(\phi,X\right)\equiv
\left(1-2D\left(\phi,X\right)X\right)^{-1} =
\left(1-1/a\right)^{-1}\left(1+2B\left(\phi\right)
X\right),\quad 0<a<1,
\end{equation}
see eq.~(\ref{eq:condjacx}) with
$$
B(\phi)=\frac{\psi^2}{A\left(\psi/\sqrt{\left\lvert\alpha\right\rvert}\right)},\quad \psi=\sqrt{\frac{-2\alpha}{\beta}}\mathrm{e}^{-\phi}.
$$
Since $\tilde{X}$ is a second-order polynomial in $X$, one gets two
possible solutions for $X$ given by
\begin{equation}
X = \frac{-1}{4B\left(\phi\right)}\left(1\pm
S\left(\phi,\tilde{X}\right)\right),\quad
S\left(\phi,\tilde{X}\right)\equiv
\sqrt{1+8B\left(\phi\right)\left(1-\frac{1}{a}\right)\tilde{X}}
\label{eq:xxtilde}
\end{equation}
Depending on which sign is chosen ($+$ or $-$), one is led to two
distinct disformed actions, $S_+$ and $S_-$ respectively. One must therefore identify which variational principle is
solved by the disformed metric (\ref{eq:wormholesolution}-\ref{eq:wormholesolution_bisbis}). To this aim, one has to analyze the situation on-shell
where
\begin{equation}
S\left(\phi,\tilde{X}\right) = \left\lvert
s\left(r\right)\right\rvert,\quad s\left(r\right)\equiv 
1-2B\left(\phi\right)\frac{f\left(r\right)}{r^2},\quad
\phi=\ln\left(\frac{\sqrt{-2\alpha/\beta}}{r}\right). \label{eq:Ss}
\end{equation}
This in turn implies that
\begin{equation}
\frac{-f\left(r\right)}{2r^2} =
\frac{-1}{4B\left(\phi\right)}\left(1\pm\left\lvert
s\left(r\right)\right\rvert\right)
\end{equation}
and, this is consistent only by choosing the $+$ sign when
$s\left(r\right)\leq 0$, and the $-$ sign when $s\left(r\right)\geq 0$. As a consequence, the disformed metric solves the
equations of motion of $S_+$ (resp. of $S_-$) if and only if
$s\left(r\right)\leq 0$ (resp. if $s\left(r\right)\geq 0$). In
particular, it will be problematic to define an action principle for
the disformed theory if the function $s\left(r\right)$ has a nonconstant sign. Note that $s\left(r\right)$ changes sign precisely at the singular radius $r_*$ identified in (\ref{eq:condjac}), thus, we retrieve the necessity of hiding $r_*$ below the wormhole throat. This is for instance ensured by our choice (\ref{eq:ourA}), for which $s\left(r\right)<0$ in the whole
physical spacetime, and hence a well-defined action principle is
shown to exist. The corresponding beyond Horndeski theory is given
by (for readability, we write coefficients as functions of variables
$\left(y,x\right)$, where $y$ stands for $\phi$ and $x$ for
$\tilde{X}$):
\begin{align*}
\tilde{F_5}\left(y,x\right) ={}& \frac{2 (a-1) \alpha
\sqrt{-\frac{x B(y)} {S\left(y,x\right)+1}} \left(a
S\left(y,x\right)+4 (a-1) x B(y)-2 S\left(y,x\right)+a\right)}
{3 a x^2 S\left(y,x\right) \left(a \left(S\left(y,x\right)-1\right)-4 (a-1) x B(y)\right)},\\
\tilde{G_5}\left(y,x\right) ={}& \frac{2 \alpha  \ln
\left(\frac{S\left(y,x\right)+1}{4 B(y)}\right)} {\sqrt{-\frac{x
B(y)}{S\left(y,x\right)+1}}}+\frac{8 \alpha
\sqrt{S\left(y,x\right)-1}
\arctan\left(\frac{\sqrt{S\left(y,x\right)-1}}{\sqrt{2}}\right)-4
\sqrt{2}
 \alpha  \ln \left(\frac{S\left(y,x\right)+1}{4 B(y)}\right)}{\sqrt{\frac{a-a S\left(y,x\right)}{a-1}}},\\
\tilde{G_4}\left(y,x\right) ={}&\frac{1}{B(y) \sqrt{-\frac{x
B(y)}{S\left(y,x\right)+1}} \sqrt{\frac{a-a S\left(y,x\right)}{a-1}}
\left(a \left(-S\left(y,x\right)\right)+8 (a-1) x
B(y)+a\right)}\left\lbrace 4 \alpha  x B'(y) \left((a-1) x
B(y)\right.\right.\\ & \left.\left. \left(8 \sqrt{2} \sqrt{-\frac{x
B(y)}{S\left(y,x\right)+1}}-\left(\sqrt{\frac{a-a
S\left(y,x\right)}{a-1}}-2 \sqrt{2} \sqrt{-\frac{x
B(y)}{S\left(y,x\right)+1}}\right) \ln
\left(\frac{S\left(y,x\right)+1}{4
B(y)}\right)\right)-\right.\right.\\ & \left.\left. \sqrt{2} a
\left(S\left(y,x\right)-1\right) \sqrt{-\frac{x
B(y)}{S\left(y,x\right)+1}}\right) \right\rbrace+2 \sqrt{-\frac{x
B(y)}{S\left(y,x\right)+1}} \left(-\frac{\alpha
\left(S\left(y,x\right)+1\right)}{B(y)}-\beta  \mathrm{e}^{2 y}+1\right),
\end{align*}
and where the expressions for $\tilde{G_2}$, $\tilde{G_3}$ and
$\tilde{F_4}$ are too cumbersome to report.


\begin{thebibliography}{99}


\bibitem{Morris:1988cz}
M.~S.~Morris and K.~S.~Thorne,
Am. J. Phys. \textbf{56} (1988), 395-412
doi:10.1119/1.15620


\bibitem{LIGOScientific:2017vwq}
B.~P.~Abbott \textit{et al.} [LIGO Scientific and Virgo],
Phys. Rev. Lett. \textbf{119}, no.16, 161101 (2017)
doi:10.1103/PhysRevLett.119.161101
[arXiv:1710.05832 [gr-qc]].


\bibitem{LIGOScientific:2020zkf}
R.~Abbott \textit{et al.} [LIGO Scientific and Virgo],
Astrophys. J. Lett. \textbf{896}, no.2, L44 (2020)
doi:10.3847/2041-8213/ab960f
[arXiv:2006.12611 [astro-ph.HE]].


\bibitem{LIGOScientific:2021qlt}
R.~Abbott \textit{et al.} [LIGO Scientific, KAGRA and VIRGO],
Astrophys. J. Lett. \textbf{915}, no.1, L5 (2021)
doi:10.3847/2041-8213/ac082e
[arXiv:2106.15163 [astro-ph.HE]].


\bibitem{Charmousis:2021npl}
C.~Charmousis, A.~Leh\'ebel, E.~Smyrniotis and N.~Stergioulas,
JCAP \textbf{02}, no.02, 033 (2022)
doi:10.1088/1475-7516/2022/02/033 [arXiv:2109.01149 [gr-qc]].


\bibitem{Horndeski:1974wa}
G.~W.~Horndeski,
Int. J. Theor. Phys. \textbf{10}, 363-384 (1974)
doi:10.1007/BF01807638


\bibitem{Babichev:2013cya}
E.~Babichev and C.~Charmousis,
JHEP \textbf{08}, 106 (2014)
doi:10.1007/JHEP08(2014)106
[arXiv:1312.3204 [gr-qc]].
T.~Kobayashi and N.~Tanahashi,
PTEP \textbf{2014}, 073E02 (2014)
doi:10.1093/ptep/ptu096
[arXiv:1403.4364 [gr-qc]].
C.~Charmousis and D.~Iosifidis,
J. Phys. Conf. Ser. \textbf{600} (2015), 012003
doi:10.1088/1742-6596/600/1/012003
[arXiv:1501.05167 [gr-qc]].
M.~Minamitsuji and J.~Edholm,
Phys. Rev. D \textbf{100}, no.4, 044053 (2019)
doi:10.1103/PhysRevD.100.044053
[arXiv:1907.02072 [gr-qc]].


\bibitem{BenAchour:2019fdf}
J.~Ben Achour, H.~Liu and S.~Mukohyama,
JCAP \textbf{02}, 023 (2020)
doi:10.1088/1475-7516/2020/02/023
[arXiv:1910.11017 [gr-qc]].
K.~Takahashi and H.~Motohashi,
JCAP \textbf{06} (2020), 034
doi:10.1088/1475-7516/2020/06/034
[arXiv:2004.03883 [gr-qc]].


\bibitem{Babichev:2017guv}
  E.~Babichev, C.~Charmousis and A.~Leh\'ebel,
  JCAP {\bf 1704} (2017) 027, doi: 10.1088/1475-7516/2017/04/027;
E.~Babichev and A.~Leh\'ebel,
JCAP \textbf{12} (2018), 027, doi: 10.1088/1475-7516/2018/12/027;
J.~Chagoya and G.~Tasinato,
  JCAP {\bf 1808}, 006 (2018), doi: 10.1088/1475-7516/2018/08/006;
T.~Kobayashi and T.~Hiramatsu,
Phys. Rev. D \textbf{97} (2018) no.10, 104012, doi: 10.1103/PhysRevD.97.104012;
A.~Leh\'ebel, E.~Babichev and C.~Charmousis,
JCAP \textbf{07} (2017), 037
doi:10.1088/1475-7516/2017/07/037
[arXiv:1706.04989 [gr-qc]].
M.~Minamitsuji and J.~Edholm,
  Phys.\ Rev.\ D {\bf 101}, no. 4, 044034 (2020), doi: 10.1103/PhysRevD.101.044034
A.~Bakopoulos, C.~Charmousis, P.~Kanti and N.~Lecoeur,
[arXiv:2203.14595 [gr-qc]].


\bibitem{Bocharova:1970skc}
N.~M.~Bocharova, K.~A.~Bronnikov and V.~N.~Melnikov,
Vestn.Mosk.Univ.Ser.III Fiz.Astron. (1970) 6, 706-709


\bibitem{Bekenstein:1974sf}
J.~D.~Bekenstein,
Annals Phys. \textbf{82}, 535-547 (1974)
doi:10.1016/0003-4916(74)90124-9


\bibitem{Klimcik:1993cia}
C.~Klimcik,
J. Math. Phys. \textbf{34}, 1914-1926 (1993)
doi:10.1063/1.530146


\bibitem{Martinez:2002ru}
C.~Martinez, R.~Troncoso and J.~Zanelli,
Phys. Rev. D \textbf{67}, 024008 (2003)
doi:10.1103/PhysRevD.67.024008
[arXiv:hep-th/0205319 [hep-th]].


\bibitem{Martinez:2005di}
C.~Martinez, J.~P.~Staforelli and R.~Troncoso,
Phys. Rev. D \textbf{74}, 044028 (2006)
doi:10.1103/PhysRevD.74.044028
[arXiv:hep-th/0512022 [hep-th]].


\bibitem{Fernandes:2021dsb}
P.~G.~S.~Fernandes,
Phys. Rev. D \textbf{103}, no.10, 104065 (2021)
doi:10.1103/PhysRevD.103.104065
[arXiv:2105.04687 [gr-qc]].


\bibitem{Kiritsis}
C.~Charmousis, B.~Gouteraux and E.~Kiritsis,
JHEP \textbf{09} (2012), 011
doi:10.1007/JHEP09(2012)011
[arXiv:1206.1499 [hep-th]].


\bibitem{Dotti:2005rc}
G.~Dotti and R.~J.~Gleiser,
Phys. Lett. B \textbf{627} (2005), 174-179
doi:10.1016/j.physletb.2005.08.110
[arXiv:hep-th/0508118 [hep-th]].


\bibitem{Bogdanos}
C.~Bogdanos, C.~Charmousis, B.~Gouteraux and R.~Zegers,
JHEP \textbf{10} (2009), 037
doi:10.1088/1126-6708/2009/10/037
[arXiv:0906.4953 [hep-th]].


\bibitem{Glavan}
D.~Glavan and C.~Lin,
Phys. Rev. Lett. \textbf{124} (2020) no.8, 081301
doi:10.1103/PhysRevLett.124.081301
[arXiv:1905.03601 [gr-qc]].


\bibitem{Zumalacarregui:2013pma}
M.~Zumalac\'arregui and J.~Garc\'\i{}a-Bellido,
Phys. Rev. D \textbf{89} (2014), 064046
doi:10.1103/PhysRevD.89.064046
[arXiv:1308.4685 [gr-qc]].


\bibitem{BenAchour:2016cay}
J.~Ben Achour, D.~Langlois and K.~Noui,
Phys. Rev. D \textbf{93}, no.12, 124005 (2016)
doi:10.1103/PhysRevD.93.124005
[arXiv:1602.08398 [gr-qc]].


\bibitem{Anson:2020trg}
T.~Anson, E.~Babichev, C.~Charmousis and M.~Hassaine,
JHEP \textbf{01}, 018 (2021)
doi:10.1007/JHEP01(2021)018
[arXiv:2006.06461 [gr-qc]].


\bibitem{Charmousis:2019vnf}
C.~Charmousis, M.~Crisostomi, R.~Gregory and N.~Stergioulas,
Phys. Rev. D \textbf{100} (2019) no.8, 084020
doi:10.1103/PhysRevD.100.084020
[arXiv:1903.05519 [hep-th]].


\bibitem{Anson:2021yli}
T.~Anson, E.~Babichev and C.~Charmousis,
Phys. Rev. D \textbf{103} (2021) no.12, 124035
doi:10.1103/PhysRevD.103.124035
[arXiv:2103.05490 [gr-qc]].


\bibitem{Bakopoulos:2021liw}
A.~Bakopoulos, C.~Charmousis and P.~Kanti,
JCAP \textbf{05} (2022) no.05, 022
doi:10.1088/1475-7516/2022/05/022
[arXiv:2111.09857 [gr-qc]].


\bibitem{Faraoni:2021gdl}
V.~Faraoni and A.~Leblanc,
JCAP \textbf{08} (2021), 037
doi:10.1088/1475-7516/2021/08/037
[arXiv:2107.03456 [gr-qc]].


\bibitem{Chatzifotis:2021hpg}
N.~Chatzifotis, E.~Papantonopoulos and C.~Vlachos,
Phys. Rev. D \textbf{105} (2022) no.6, 064025
doi:10.1103/PhysRevD.105.064025
[arXiv:2111.08773 [gr-qc]].


\bibitem{Kanti:2011jz}
P.~Kanti, B.~Kleihaus and J.~Kunz,
Phys. Rev. Lett. \textbf{107} (2011), 271101
doi:10.1103/PhysRevLett.107.271101
[arXiv:1108.3003 [gr-qc]].
P.~Kanti, B.~Kleihaus and J.~Kunz,
Phys. Rev. D \textbf{85} (2012), 044007
doi:10.1103/PhysRevD.85.044007
[arXiv:1111.4049 [hep-th]].
G.~Antoniou, A.~Bakopoulos, P.~Kanti, B.~Kleihaus and J.~Kunz,
Phys. Rev. D \textbf{101} (2020) no.2, 024033
doi:10.1103/PhysRevD.101.024033
[arXiv:1904.13091 [hep-th]].


\bibitem{Lovelock:1971yv}
D.~Lovelock,
J. Math. Phys. \textbf{12} (1971), 498-501 doi:10.1063/1.1665613


\bibitem{Charmousis:2008kc}
C.~Charmousis,
Lect. Notes Phys. \textbf{769} (2009), 299-346
doi:10.1007/978-3-540-88460-6\_8 [arXiv:0805.0568 [gr-qc]].


\bibitem{Lu:2020iav}
H.~Lu and Y.~Pang,
Phys. Lett. B \textbf{809} (2020), 135717
doi:10.1016/j.physletb.2020.135717
[arXiv:2003.11552 [gr-qc]].
R.~A.~Hennigar, D.~Kubiz\v{n}\'ak, R.~B.~Mann and C.~Pollack,
JHEP \textbf{07} (2020), 027
doi:10.1007/JHEP07(2020)027
[arXiv:2004.09472 [gr-qc]].


\bibitem{Fernandes:2022zrq}
P.~G.~S.~Fernandes, P.~Carrilho, T.~Clifton and D.~J.~Mulryne,
Class. Quant. Grav. \textbf{39} (2022) no.6, 063001
doi:10.1088/1361-6382/ac500a
[arXiv:2202.13908 [gr-qc]].


\bibitem{Hui:2012qt}
L.~Hui and A.~Nicolis,
Phys. Rev. Lett. \textbf{110} (2013), 241104
doi:10.1103/PhysRevLett.110.241104 [arXiv:1202.1296 [hep-th]].
T.~P.~Sotiriou and S.~Y.~Zhou,
Phys. Rev. D \textbf{90} (2014), 124063
doi:10.1103/PhysRevD.90.124063
[arXiv:1408.1698 [gr-qc]].
E.~Babichev, C.~Charmousis and A.~Leh\'ebel,
Class. Quant. Grav. \textbf{33} (2016) no.15, 154002
doi:10.1088/0264-9381/33/15/154002
[arXiv:1604.06402 [gr-qc]].


\bibitem{Myers:1988ze}
R.~C.~Myers and J.~Z.~Simon,
Phys. Rev. D \textbf{38} (1988), 2434-2444
doi:10.1103/PhysRevD.38.2434


\bibitem{Clunan:2004tb}
T.~Clunan, S.~F.~Ross and D.~J.~Smith,
Class. Quant. Grav. \textbf{21} (2004), 3447-3458
doi:10.1088/0264-9381/21/14/009 [arXiv:gr-qc/0402044 [gr-qc]].


\bibitem{Iyer:1994ys}
V.~Iyer and R.~M.~Wald,
Phys. Rev. D \textbf{50}, 846-864 (1994) doi:10.1103/PhysRevD.50.846
[arXiv:gr-qc/9403028 [gr-qc]].


\bibitem{Ayon-Beato:2005yoq}
E.~Ayon-Beato, C.~Martinez, R.~Troncoso and J.~Zanelli,
Phys. Rev. D \textbf{71} (2005), 104037
doi:10.1103/PhysRevD.71.104037 [arXiv:hep-th/0505086 [hep-th]].


\bibitem{Ayon-Beato:2004nzi}
E.~Ayon-Beato, C.~Martinez and J.~Zanelli,
Gen. Rel. Grav. \textbf{38}, 145-152 (2006)
doi:10.1007/s10714-005-0213-x [arXiv:hep-th/0403228 [hep-th]].


\bibitem{Charmousis:2014mia}
C.~Charmousis,
Lect. Notes Phys. \textbf{892}, 25-56 (2015)
doi:10.1007/978-3-319-10070-8\_2 [arXiv:1405.1612 [gr-qc]].
C.~Charmousis and D.~Iosifidis,
J. Phys. Conf. Ser. \textbf{600} (2015), 012003
doi:10.1088/1742-6596/600/1/012003
[arXiv:1501.05167 [gr-qc]].


\bibitem{Hartle:1967he}
J.~B.~Hartle,
Astrophys. J. \textbf{150} (1967), 1005-1029 doi:10.1086/149400


\bibitem{Hartle:1968si}
J.~B.~Hartle and K.~S.~Thorne,
Astrophys. J. \textbf{153} (1968), 807 doi:10.1086/149707


\bibitem{Bakopoulos:2022csr}
A.~Bakopoulos, C.~Charmousis, P.~Kanti and N.~Lecoeur,
[arXiv:2203.14595 [gr-qc]].


\bibitem{Bettoni:2013diz}
D.~Bettoni and S.~Liberati,
Phys. Rev. D \textbf{88} (2013), 084020
doi:10.1103/PhysRevD.88.084020 [arXiv:1306.6724 [gr-qc]].


\bibitem{Gleyzes:2014qga}
J.~Gleyzes, D.~Langlois, F.~Piazza and F.~Vernizzi,
JCAP \textbf{02} (2015), 018 doi:10.1088/1475-7516/2015/02/018
[arXiv:1408.1952 [astro-ph.CO]].


\bibitem{Barriola:1989hx}
M.~Barriola and A.~Vilenkin,
Phys. Rev. Lett. \textbf{63} (1989), 341
doi:10.1103/PhysRevLett.63.341


\bibitem{Damour:2007ap}
T.~Damour and S.~N.~Solodukhin,
Phys. Rev. D \textbf{76} (2007), 024016
doi:10.1103/PhysRevD.76.024016
[arXiv:0704.2667 [gr-qc]].


\cite{Bettoni:2013diz}
\bibitem{Bettoni:2013diz}
D.~Bettoni and S.~Liberati,
Phys. Rev. D \textbf{88} (2013), 084020
doi:10.1103/PhysRevD.88.084020
[arXiv:1306.6724 [gr-qc]].


\bibitem{Evseev:2017jek}
O.~A.~Evseev and O.~I.~Melichev,
Phys. Rev. D \textbf{97} (2018) no.12, 124040
doi:10.1103/PhysRevD.97.124040
[arXiv:1711.04152 [gr-qc]].


\bibitem{Mironov:2018uou}
S.~Mironov, V.~Rubakov and V.~Volkova,
Class. Quant. Grav. \textbf{36} (2019) no.13, 135008
doi:10.1088/1361-6382/ab2574
[arXiv:1812.07022 [hep-th]].


\bibitem{Franciolini:2018aad}
G.~Franciolini, L.~Hui, R.~Penco, L.~Santoni and E.~Trincherini,
JHEP \textbf{01} (2019), 221
doi:10.1007/JHEP01(2019)221
[arXiv:1811.05481 [hep-th]].


\bibitem{Fernandes:2020nbq}
P.~G.~S.~Fernandes, P.~Carrilho, T.~Clifton and D.~J.~Mulryne,
Phys. Rev. D \textbf{102} (2020) no.2, 024025
doi:10.1103/PhysRevD.102.024025
[arXiv:2004.08362 [gr-qc]].

\end{thebibliography}
\end{document}